\documentclass[onecolumn]{els-mrw} 

\usepackage{amsmath,amssymb,amsfonts,amsthm,makeidx,graphicx}
\usepackage{txfonts}
\usepackage{helvet}
\usepackage{comment}

\usepackage{fontspec}

\usepackage{color}

\newcommand{\width}{\xi}

\newcommand{\E}[0]{\mathcal{E}}

\newcommand{\ket}[1]{\left| #1 \right\rangle}
\newcommand{\bra}[1]{\left\langle #1 \right|}
\newcommand{\braket}[2]{\left\langle #1| #2 \right\rangle}
\newcommand{\bracket}[2]{\langle #1 | #2  \rangle}

\newcommand{\eff}[0]{\mathrm{eff}}

\def\n{\noindent}
\def\b{\bigskip}

\def\bn{\bigskip\noindent}

\def\I{{\cal I}}
\def\F{{\cal F}}

\def\F{{\cal F}}

\def\H{{\cal H}}
\def\M{{\cal M}}

\def\bC{{\Bbb C}}

\def\bR{{\Bbb R}}

\def\p{\partial}
\def\h{\hspace}

\def\eps{\varepsilon}

\makeatletter
\let\Hy@backout\@gobble
\makeatother

\usepackage{color}

\begin{document}

\chapter{Chaos and Quantum Tunneling}
\label{chap1}

\author[1]{Akira Shudo}%
%

\address[1]{\orgname{Tokyo Metropolitan University}, \orgdiv{Department of Physics}, \orgaddress{Minami-Osawa, Hachioji, Tokyo 192-0397, Japan}}

\articletag{Chapter Article tagline: Sept. 1, 2025}

\maketitle


\begin{glossary}[Keywords]
Chaos, Dynamical Tunneling, Chaos-assisted Tunneling, Chaotic Tunneling, Resonance-assisted Tunneling, Instanton, Complex Semiclassical Method, Julia Set, Stable and Unstable Manifolds

\end{glossary}

\begin{abstract}[Abstract]
In generic Hamiltonian systems that are neither completely integrable nor fully chaotic, phase space consists of a mixture of regular and chaotic components.
In classical dynamics, transitions between different invariant sets in phase space are strictly forbidden, and these sets act as dynamical barriers to one another.
In quantum mechanics, in contrast, wave effects allow transitions through such dynamical barriers.
This process, known as dynamical tunneling, refers to penetration through dynamical barriers in phase space and was first recognized in the early 1980s.
Since then, various aspects of dynamical tunneling have been elucidated, significantly advancing our understanding of such a novel quantum phenomenon.
In this article, we provide an overview of several phenomenological perspectives of dynamical tunneling, including chaos-assisted and resonance-assisted tunneling, and also introduce approaches based on classical mechanics extended into the complex domain.
In particular, we seek to clarify what is meant by the common claim that ``chaos leads to an enhancement of the tunneling probability”, which is often made when dynamical tunneling is dressed.  
We discuss what regime this refers to and, if such an enhancement occurs, what its likely origin is.

\end{abstract}

\section{Introduction}
\label{sec:introduction}

Chaos is a phenomenon that arises in classical mechanics, whereas quantum tunneling is intrinsically quantum and admits no classical analogue. 
Hence, there appears to be no direct link between chaos and quantum tunneling. Nevertheless, precisely for this reason, the question of how non-integrability, and the resulting classical chaos, influences tunneling in the corresponding quantum system remains a fascinating problem in quantum physics.
In this article, we explore how these two seemingly disparate phenomena are connected and how chaos influences quantum tunneling, surveying the studies carried out to date and highlighting the open questions to be examined in the future.

Studies on quantum tunneling, in both one- and multi-dimensional settings, have long been carried out regardless of whether the underlying classical dynamics exhibits chaos.
In the conventional setting, quantum tunneling refers to the phenomenon in which, for a system with a potential barrier, a particle with energy below the barrier height nevertheless penetrates the barrier due to quantum-mechanical effects. However, in systems with two or more degrees of freedom, the motion is no longer constrained solely by energy barriers: additional (local) integrals of motion impose dynamical restrictions, leading to qualitatively new situations. 

\textit{Dynamical~tunneling} refers to classically forbidden transitions that proceed through dynamical, rather than energetic, barriers that are formed in classical phase space~\cite{davis1981quantum,Creagh98,keshavamurthy2011dynamical}. 
Dynamical tunneling can be regarded as a generalization of barrier tunneling; 
however, as discussed in detail in this article, 
the presence of chaos renders the relation between dynamical and energy-barrier tunneling far more subtle than is commonly appreciated. One then realizes that even providing a precise definition of dynamical tunneling is more difficult than it sounds. 
Quantum tunneling becomes especially  challenging when the underlying classical dynamics is non-integrable.
Even in integrable systems, tunneling with two or more degrees of freedom already departs from the textbook one-dimensional paradigm and is of substantial interest~\cite{creagh1994tunnelling}. 
Here, however, our emphasis is on the fundamental distinctions between tunneling in integrable and non-integrable systems.

Here, we introduce ideas that provide a phenomenological perspective on quantum tunneling in non-integrable systems, including \textit{chaos-assisted tunneling} (CAT)~\cite{Bohigas93,tomsovic1994chaos} and \textit{resonance-assisted tunneling} (RAT)~\cite{brodier2001,brodier2002}, both of which are nowadays widely recognized as characteristic mechanisms of dynamical tunneling in mixed phase spaces. 
On the other hand, we are naturally led to ask more directly what aspects of chaos are represented by such phenomena.
For this purpose, a reasonable option would be to use trajectories in complex space.
In fact, attempts to describe tunneling in terms of trajectories in complex space date back to well before the advent of discussions on dynamical tunneling~\cite{maslov2012complex,landau1932theory,zener1932non,stuckelberg1932theory,langer1937connection,perelomov1966ionization,balian1970distribution,george1972complex,berry1972semiclassical,mclaughlin1972complex,miller1974quantum,knoll1976semiclassical,callan1977fate,coleman1977fate,voros1983}.
In the stationary-phase, or saddle-point, approximation to the path integral for quantum transitions, the contributing trajectories are real in classically allowed processes, whereas in classically forbidden processes such as tunneling, the relevant trajectories are complex and are obtained by integrating the equations of motion in complex time or complex coordinates.

Dynamical tunneling has likewise motivated a variety of studies aiming to interpret tunneling processes in terms of complex trajectories and complex phase space~\cite{wilkinson1986tunnelling,shudo1995,doron1995semiclassical,creagh1996complex,takahashi2000complex,levkov2007complex,backer2008regular}.
So far, however, the connection between phenomenological approaches and complex-trajectory-based analyses has not been discussed in a systematic manner. 
The primary aim of the present article is therefore to relate these two different approaches to quantum tunneling in non-integrable systems. We expect this comparison to provide a useful resource for readers interested in dynamical tunneling.

The outline of this article is as follows. 
In Sec.~\ref{sec:Role_chaos}, we explain a basic idea of CAT and introduce an attempt to interpret it in terms of a hybrid semiclassical approach developed in Ref.~\cite{doron1995semiclassical,frischat1998dynamical}.  
The emergence of chaos is one of the most conspicuous hallmarks of non-integrability, and it is therefore natural to ask how the presence of chaos influences quantum tunneling. 
Dynamical tunneling, as proposed by Davis and Heller~\cite{davis1981quantum}, typically manifests as tunneling doublets formed by (symmetry-related) local modes supported by regular regions in phase space. 
In mixed systems, these tunneling doublets are expected to be influenced by a chaotic sea that lies between the regular islands.
As a system parameter varies, the energies of the tunneling doublet formed by the local modes and the energies of the eigenstates supported by the intervening chaotic sea approach each other and become nearly degenerate. 
This causes the tunneling splitting to grow {\it resonantly}. 
The increase in the tunneling splitting implies that the transition probability for tunneling between the regular regions is enhanced. 
The scattering matrix formulation provides a clue for understanding the role of chaos in such a circumstance based on complex classical dynamics~\cite{doron1995semiclassical,frischat1998dynamical}.

In Sec.~\ref{sec:Role_resonance}, we discuss how nonlinear resonances in phase space manifest themselves in the tunneling process.  
The correspondence between avoided crossings of energy levels and nonlinear resonances in the underlying classical dynamics has long been a subject of investigation, especially in the field of chemical physics~\cite{ramaswamy1981perturbative,uzer1983uniform,ramachandran1993influence,roberts1993correspondence,uzer1991theories}. 
In the context of dynamical tunneling, especially in connection with the resonant spikes in the tunneling splitting observed near avoided crossings, 
it was found that a resonant enhancement of the tunneling splitting can also occur when a third state associated with a classical nonlinear resonance undergoes an avoided crossing with the tunneling doublet supported by a regular region~\cite{bonci1998tunneling}.  
This observation suggests that the resonant enhancement of the tunneling splitting is not due exclusively to the CAT mechanism.
Along this line, Brodier et al. proposed a hybrid classical-quantum scheme to incorporate the influence of nonlinear resonances into tunneling~\cite{brodier2001,brodier2002}. 
Subsequently, the proposed method has been tested for validity across various systems, with incremental refinements~\cite{eltschka2005resonance,mouchet2006influence,wimberger2006resonance,keshavamurthy2005resonance,lock2010,Schlagheck11}.
Here, we briefly outline the practical computational scheme of RAT and illustrate it with applications to two-dimensional symplectic maps. 
Through such a demonstration, we undertake a detailed examination of the assumptions underlying the RAT approach and the interpretation of the results.
In spite of the prediction from the instanton-based calculation~\cite{simon1983,simon1984}, 
one often encounters a characteristic step-like dependence of the tunneling splitting on $1/\hbar$, meaning the deviation from a simple exponential law~\cite{brodier2001,brodier2002,Mouchet03,le2013semiclassical,lock2010,hanada2015,hanada2023dynamical}. 
This may provide a crucial difference from quantum tunneling in integrable systems, and identifying the origin of such deviations is therefore one of the major challenges. 
To this end, we introduce an approach based on \textit{ultra-near-integrable systems}, recently proposed in Refs.~\cite{iijima2022quantum,koda2023ergodicity}, and argue that \textit{quantum resonances}, rather than classical resonances, play a crucial role. 

In Sec.~\ref{sec:Complex_path}, we introduce complex-path approaches to dynamical tunneling.
First, we provide an overview of the complex-trajectory approaches that have been developed so far, mainly for continuous-time systems.
We then develop arguments based on fully complex semiclassical analysis in the time domain using discrete maps. 
It will be emphasized that 
the complex trajectories that contribute to tunneling differ fundamentally from those in one-dimensional systems, not only in their number but also in their qualitative characteristics.
We point out that the candidate complex trajectories in non-integrable maps lie on the {\it Julia set} in the complex domain~\cite{shudo2002,shudo2009a,shudo2009b}. 
Among the family of candidate trajectories contained in the Julia set, finding complex trajectories that give the dominant contribution becomes the next task~\cite{koda2022complexified}. 
To this end, mathematical results on multi-dimensional complex dynamical systems offer essential insights~\cite{bedford1991,bedford1991b,bedford1992a,bedford1992b}.

In Sec.~\ref{sec:conclutions}, we provide a summary and outline future issues to be addressed. 
Although this article focuses mainly on theoretical aspects of chaos and quantum tunneling, 
we should mention that dynamical tunneling has been observed in a variety of experimental platforms,
including cold-atom experiments, superconducting billiards, microwave billiards, and optical microcavities~\cite{dembowski2000first,hensinger2001dynamical,steck2001observation,hofferbert2005experimental,backer2008dynamical,shinohara2010chaos,kim2013chaos,gehler2015experimental,arnal2020chaos}.
We also do not enter into the issue of how dynamical tunneling is involved in chemical reactions.
Intramolecular vibrational energy redistribution (IVR) refers to the flow of energy from a specific vibrational mode to other modes within a molecule, and dynamical tunneling plays a key role in such a process.
The rates and pathways of IVR are governed by the interplay between classical diffusion in phase space and dynamical tunneling.
Moreover, in polyatomic molecules, Arnold diffusion, which arises in systems with three or more degrees of freedom, can also contribute to IVR, making it necessary to carefully investigate dynamical tunneling in high-dimensional settings. 
For comprehensive reviews of these topics, Refs.~\cite{keshavamurthy2005dynamical,keshavamurthy2007dynamical} are highly recommended.

\section{Role of chaos in quantum tunneling}
\label{sec:Role_chaos}

\subsection{Avoided crossing and resonant spikes in tunneling splitting}
\label{sec:spikes}%

First, let us consider a situation in which chaos is relatively well developed around the regular region in classical phase space.
Lin and Ballentine investigated a periodically driven system whose phase space contains two regular regions surrounded by a chaotic sea. They found that coherent oscillatory tunneling occurs and that the tunneling rate is several orders of magnitude larger than that of ordinary tunneling without the driving force~\cite{lin1990quantum,lin1992quantum}.  
Utermann and co-workers, on the other hand, calculated the eigenstates (Floquet states) of a periodically driven system that realizes a mixed phase space, as well as evaluated the tunneling rate when a wave packet initially localized on one regular region penetrates into the other regular region. 
They showed that the tunneling rate is governed by the overlap between the regular islands and the chaotic region that separates them, not by the overlap between regular islands supporting a doublet~\cite{utermann1994tunneling}.

The term chaos-assisted tunneling (CAT) was introduced by Bohigas, Tomsovic, and Ullmo in Ref.~\cite{Bohigas93}.  
CAT revealed the role of chaos in dynamical tunneling proposed by Davis and Heller~\cite{davis1981quantum}.
It becomes relevant when one moves from a nearly integrable regime to a regime in which the chaotic component of phase space has grown substantially. 
At the same time, chaos is not assumed to occupy the entire phase space; regular regions supporting tunneling doublets are assumed to persist with sufficient measure.
To see the influence of classical chaos on quantum tunneling, 
they employed the autonomous two-degree-of-freedom system with a quartic symmetric polynomial,
focusing mainly on the following aspects:  
(1) the response of the tunneling splitting to variations in system parameters, and  
(2) the statistical distribution of the tunneling splittings.

\begin{figure}[hbt]
\centering
\includegraphics[width=1.1\linewidth, trim=0 90mm 0 95mm, clip]{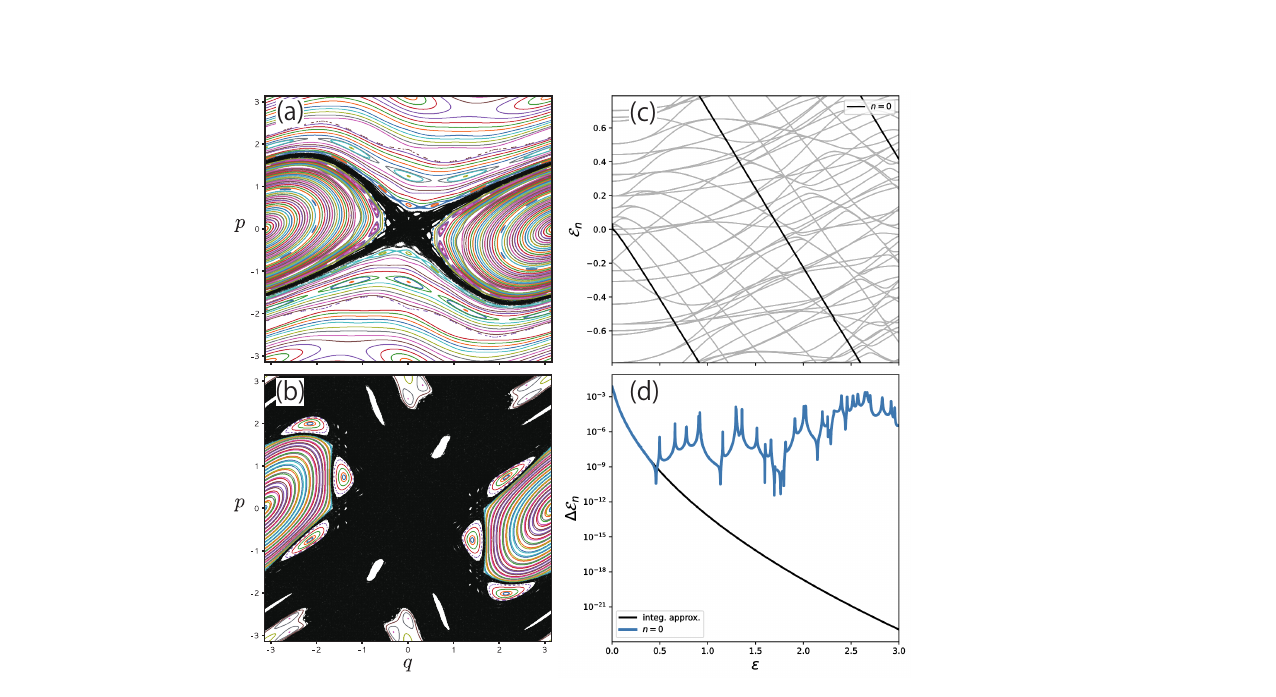}
\caption{\label{fig:map_splitting}
Phase-space portraits of the classical map $f$ for (a) $\varepsilon=0.7$ and (b) $\varepsilon=1.5$, respectively. 
(c) The black curves indicate the energies of the ground-state doublet, while the gray ones show the quasi-energy spectrum for $1/\hbar=3.979$.
(d) The blue and black curves illustrate the tunneling splitting of the ground-state doublet shown in (c) and the corresponding splitting for the BCH Hamiltonian, respectively.
For the BCH Hamiltonian, see Sec.~\ref{sec:Verification_of_RAT}.  
Reproduced from an unpublished calculation by Yasutaka Hanada.
}
\end{figure}

Let us now examine each of these aspects in detail. 
For consistency with the subsequent demonstrations in this article, we use the kicked rotor model here as an illustrative example, rather than the autonomous Hamiltonian: 
\begin{equation}\label{eq:kicked-ham}
 H(p,q,t) = T(p) + \epsilon V(q)\sum_{n \in \mathbb{Z}}\delta(t-n\tau), 
\end{equation}
where $\epsilon$ and $\tau$ are the strength and period of a perturbation, respectively.
The angular frequency of the perturbation is defined as $\Omega = 2\pi/\tau$.
The classical map $f$ derived from (\ref{eq:kicked-ham}) is expressed as
\begin{equation}
\label{eq:cmap}
f: \left(\begin{array}{c}q \\p\end{array}\right) 
\mapsto
\left(\begin{array}{c}
q + \tau T'(p)
\\
p - \tau  V'(q + \tau T'(p))
\end{array}\right), 
\end{equation}
which is equivalent to the second-order symplectic integrator (scheme) for the autonomous Hamiltonian 
$H(p,q) = T(p) + \epsilon V(q)$ with a time-step size $\tau$. 
In the following, we set $T(p)=p^2/2$, $V(q)=\cos q$, and $\tau=1$.
Typical phase space portraits generated by the classical map $f$ are illustrated in Figs.~\ref{fig:map_splitting}(a) and \ref{fig:map_splitting}(b). 

The quantum dynamics is described by the unitary operator
\begin{equation}
\label{eq:qmap}
 \hat{U} = e^{-\frac{i}{\hbar}\frac{\epsilon}{2}V(\hat{q})}
e^{-\frac{i}{\hbar}T(\hat{p})}
e^{-\frac{i}{\hbar}\frac{\epsilon}{2}V(\hat{q})},
\end{equation}
which is referred to as the quantum map~\cite{berry1979quantum,casati1979lecture}. 
Here, we adopt the symmetrized form of the unitary operator.
We focus on quasi-stationary states of the quantum map (\ref{eq:qmap}).
The eigenvalue equation is given as
\begin{equation}
\label{eq:qmap_eigen}
  \hat{U}\ket{\Psi_n} = u_n\ket{\Psi_n},\qquad u_n = e^{-\frac{i}{\hbar}\E_n},
\end{equation}
where $\ket{\Psi_n}$ is a quasi-eigenstate (Floquet state) and $\E_n$ is the associated quasi-eigenenergy.
In this Section, we use the above model to discuss CAT, but will continue to employ it in the subsequent Sections to explain tunneling in a nearly integrable regime. 

As seen in Fig.~\ref{fig:map_splitting}(c), the energy levels of the Floquet operator vary as the parameter $\epsilon$ increases.
This variation follows the so-called level dynamics~\cite{nakamura1986complete} and displays 
 numerous avoided crossings in general. 
Note that the system has parity and translational symmetries, so the Hilbert space can be decomposed as a direct sum of four subspaces, each characterized by a distinct set of symmetry quantum numbers.
Figure~\ref{fig:map_splitting}(d) shows the tunneling splitting for the ``ground-state" doublet.
When the parameter $\epsilon$ is sufficiently small, the tunneling splitting exhibits only smooth variation with respect to the parameter. 
On the other hand, as $\epsilon$ becomes larger, sharp spikes appear at specific values where the energy levels undergo avoided crossings.

We will not go into the origin of the phenomenon that looks like a transition here (around $\epsilon \simeq 0.33$)~\cite{hanada2023dynamical}. 
Instead, as pointed out in \cite{tomsovic1994chaos}, in regions where chaos is well developed, 
the tunneling splitting exhibits a sharp response to parameter variations.
Since the magnitude of the tunneling splitting reflects the tunneling rate between states whose supports are symmetrically located in phase space, the appearance of sharp spikes indicates an enhancement of the tunneling rate. 
These observations may lead one to conclude that {\it chaos enhances the tunneling probability}. 
However, as we discuss in detail below, sharp spikes can appear even when chaos is not yet fully developed in phase space~\cite{bonci1998tunneling}, or even when the system is integrable~\cite{le2010instantons}. 
Therefore, one cannot conclude from the appearance of spikes alone that classical chaos enhances the tunneling probability.
We would first like to clarify what is exactly meant by saying that chaos assists tunneling
and what should serve as the baseline for defining enhancement relative to an integrable system. 
This question is one of the central themes throughout the present article.
To begin with, we briefly review the work that gave rise to the term ``chaos-assisted tunneling''~\cite{Bohigas93,tomsovic1994chaos}.

\subsection{Chaos-assisted tunneling}
\label{sec:CAT}

As is well known, no closed-form semiclassical expressions are available for evaluating the tunneling splitting in chaotic systems. 
Consequently, one cannot directly determine which specific aspects of classical chaos influence the splitting and how they do so. 
For this reason, seeking a statistical characterization of tunneling splittings would be a reasonable option~\cite{tomsovic1994chaos,leyvraz1996level}. 
In Ref.~\cite{leyvraz1996level}, the statistical law obeyed by the tunneling splitting was investigated by introducing a matrix model of the following form: 
\begin{eqnarray}
H^{+}=
\begin{pmatrix}
E_{R} & v_{1}^{+} & v_{2}^{+} & \cdots\\
v_{1}^{+} & E_{1}^{+} & 0 & 0 & \cdots\\
v_{2}^{+} & 0 & E_{2}^{+} & 0 & \cdots\\
\vdots & 0 & 0 & \ddots & 
\end{pmatrix},
\qquad
H^{-}=
\begin{pmatrix}
E_{R} & v_{1}^{-} & v_{2}^{-} & \cdots\\
v_{1}^{-} & E_{1}^{-} & 0 & 0 & \cdots\\
v_{2}^{-} & 0 & E_{2}^{-} & 0 & \cdots\\
\vdots & 0 & 0 & \ddots &
\end{pmatrix}. 
\label{2.1}
\end{eqnarray}
The model represents a situation in which the quantum states, 
$ \lvert R,+\rangle $ 
and 
$ \lvert R,-\rangle $, 
supported by regular regions located symmetrically in the classical phase space, are coupled with exponentially small amplitudes 
$ v_n^{+} $ 
and 
$ v_n^{-} $ 
to chaotic states 
$ \lvert n,+\rangle $ 
and 
$ \lvert n,-\rangle $ 
($ n = 1, 2, \cdots $) lying between (or surrounding) them. 
Here, $E_R$ denotes the energy of the regular state, while $E_n^{+}$ and $E_n^{-}$ ($n = 1, 2, \cdots$) are those of the chaotic states. 
Reflecting their chaotic nature, these energies are assumed to follow independent GOE (Gaussian Orthogonal Ensemble) distributions.
Note that, as for the coupling strength $v_n^{\pm}$ between the regular and the chaotic states, 
there are approaches to evaluate it based on the so-called {\it fictitious integrable system}~\cite{backer2008regular,backer2010direct} or to evaluate the matrix element for coupling between regular and chaotic states by introducing the tile model~\cite{podolskiy2003semiclassical}. 
Here, following the argument in Ref.~\cite{tomsovic1994chaos,leyvraz1996level}, we assume that the coupling strength obeys independent Gaussian distributions (in practical calculations, these are further simplified and replaced by their mean value $v_t$). 
Moreover, although a direct tunneling coupling $\varepsilon$ should also exist between the quantum states $\lvert R,+\rangle$ and $\lvert R,-\rangle$, the paper \cite{leyvraz1996level} considered the regime $\varepsilon \ll v_n^{\pm}$, i.e., where the direct coupling is much smaller than the couplings to the chaotic states. Under this assumption, they set $\varepsilon=0$.

If we denote by $ \delta^{+} $ and $ \delta^{-} $ the shifts obtained within each symmetry subspace by diagonalizing $ H^{+} $ and $ H^{-} $, respectively, then the tunneling splitting is defined by  
\begin{eqnarray}
\delta = \lvert  \, \delta^{+} - \delta^{-} \,\rvert. 
\end{eqnarray}
When no $ E_{i}^{\pm}$ lie near $ E_{R} $, 
the standard perturbation theory predicts 
\begin{eqnarray}
\delta^{\pm} \simeq \sum_{i=1}^{N} \frac{\lvert v_{i} \rvert^{2}}{E_{R} - E_{i}^{\pm}} . 
\label{eq:3.2}
\end{eqnarray}
However, one has to handle carefully the rare but important case in which one of the chaotic levels 
$E_{i}^{\pm}$ comes very close to $E_{R}$.
In that case, one must perform an exact $ 2 \times 2 $ diagonalization for each chaotic eigenstate 
and use the expression  
\begin{eqnarray}
\delta^{\pm} = \frac{1}{2} \sum_{i=1}^{N} (E_{R} - E_{i}^{\pm})
\left( 1 - \sqrt{\,1 + \Bigl( \frac{2 v_{i}^{\pm}}{E_{R} - E_{i}^{\pm}} \Bigr)^{2} } \right). 
\label{3.1}
\end{eqnarray}
Based on this setup, Ref.~\cite{leyvraz1996level} shows that the splitting distribution $ p(\delta) $ obeys the truncated Cauchy law: 
\begin{eqnarray}
p(\delta) =
\begin{cases}
\dfrac{4 v_{t}}{\delta^{2} + 4 \pi v_{t}^{2}}, & \delta < v_{t}, \\[6pt]
0, & \delta > v_{t}. 
\end{cases}
\label{eq:2.4}
\end{eqnarray}
This theoretical prediction reproduces well the numerical results obtained for the model system~\cite{leyvraz1996level}.
It should be noted that the above treatment holds equally well both in the limit where the chaotic energy levels are completely uncorrelated (i.e., Poisson distribution) and in the limit where they are fully correlated (i.e., rigid spectrum)~\cite{leyvraz1996level}.

In general, in systems with mixed phase space, 
classical trajectories do not uniformly explore the chaotic region; rather, they become trapped for long times within certain subregions and then cross partial barriers to move elsewhere. 
As a consequence, chaotic states in the quantum system tend to localize within partially decoupled regions of phase space, which introduces correlations between the chaotic spectra of even and odd parities and modifies the distribution of splittings. 
Ref.~\cite{leyvraz1996level} also investigates tunneling splittings under such circumstances.

An important contribution related to the statistical analysis of tunneling splittings was made by Creagh and Whelan~\cite{creagh2000statistics}, who derived the statistical properties of the tunneling probability in the presence of chaos. 
In the above discussion, the tunneling couplings,  $v_n^+$ and $v_n^-$,  
were assumed to follow the Porter-Thomas distribution derived under the assumption of GOE statistics, but in Ref.~\cite{creagh2000statistics} they showed that, in fact, the distribution depends on the stability of the instanton orbit governing the tunneling process and is therefore not universal. 
They also showed that, as the instanton orbit becomes more unstable, the distribution reduces to the Porter-Thomas form.

\subsection{Scattering matrix approach}
\label{sec:Scattering}

To verify why chaos assists tunneling in systems with mixed phase space, it is necessary to investigate the mechanism of transitions between the two states supporting a tunneling doublet. 
By varying a parameter of the annular billiard, Bohigas et al. showed that transitions between the two states associated with a tunneling doublet are governed not by direct paths connecting them, but by transitions mediated through the intervening chaotic region~\cite{Bohigas93a}. 
However, in mixed systems, parameter changes simultaneously affect both the tunneling amplitude from (or into) the torus and the properties of the intermediate chaotic layer, making it difficult to disentangle which process is dominant.

The annular billiard used in \cite{Bohigas93a} is also well suited for applying the scattering map approach %
developed by Frischat and Doron~\cite{doron1995semiclassical,frischat1998dynamical}. 
It begins by expressing the wavefunction inside the annulus as a superposition of incoming and outgoing cylindrical waves, 
\begin{eqnarray}
\psi (r,\varphi)=\sum_{n=-\infty}^{\infty}\bigl[\alpha_n H_n^{(2)}(kr)+\beta_n H_n^{(1)}(kr)\bigr]e^{i n\varphi},
\end{eqnarray}
where $H_n^{(1,2)}(x)$ denote the Hankel functions of the first and second kind of order $n$, and $k$ is the wavenumber. 
The order $n$ is the angular momentum quantum number and it tends to the classical impact parameter $ L = n/k $ in the semiclassical limit. 

\begin{figure}[ht]
\centering
\includegraphics[width=0.48\linewidth]{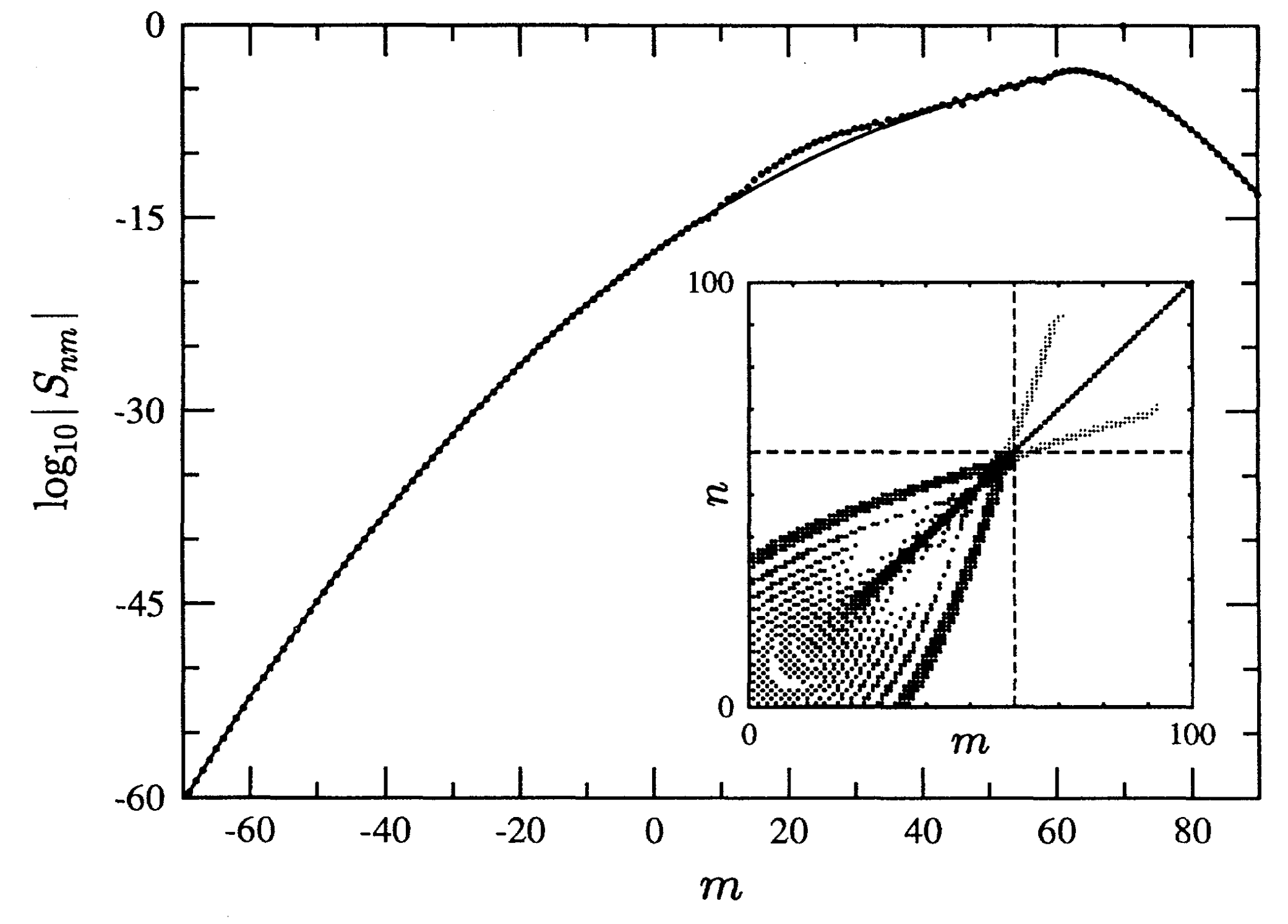}
\caption{
\label{fig:Frischat_Doron}
Tunneling amplitudes $|S_{nm}|$ for $n=70$, $a=0.4$, $\delta=0.2$, and $k=100$ as a function of $m$, calculated exactly (dots) and semiclassically (full line). The inset shows one quadrant of $|S|$, with larger values corresponding to larger dots, in arbitrary units. Off-diagonal ridges at $n,m>k(a+\delta)$ are exponentially enhanced in the plot. The dashed lines indicate $k(a+\delta)$.
Reproduced from Ref.~\cite{doron1995semiclassical}. 
}
\end{figure}

Here, we regard the annular billiard as a system composed of two scatterers. Specifically, the inner subsystem reflects an incoming wave into an outgoing one at the outer surface of the inner circle, whereas the outer subsystem scatters an outgoing wave into an incoming one at the inner surface of the outer circle.
These two systems are defined by their scattering matrices $ S^{\rm (I,O)}(k) $ and are related through the coefficient vectors $ \alpha $ and $ \beta $ by  
\begin{eqnarray}
\beta = S^{\rm (I)}(k)\,\alpha, \qquad \alpha = S^{\rm (O)}(k)\,\beta. 
\end{eqnarray}
Requiring these two relations to be simultaneously satisfied leads to the quantization condition, 
\begin{eqnarray}
\det\bigl[ S(k) - 1 \bigr] = 0, \qquad S(k) = S^{\rm (I)}(k)\,S^{\rm (O)}(k). 
\label{1}
\end{eqnarray}
Note that each time one of the eigenphases of $S(k)$ becomes an integer multiple of $2\pi$, the billiard gains an eigenvalue. 
Here $S^{\rm (O)} (k)$ is a diagonal matrix given by
\begin{eqnarray}
S^{\rm (O)}_{nm}=-\,\frac{H^{(2)}_{n}(kR)}{H^{(1)}_{n}(kR)}\,\delta_{nm}, 
\end{eqnarray}
and $S^{\rm (I)}$ is expressed using the Bessel function $J_{\,n}(x)$ as
\begin{eqnarray}
S^{\rm (I)}_{nm}
= -\sum_{\ell=-\infty}^{\infty}
J_{\,n-\ell}(k\delta)\,J_{\,m-\ell}(k\delta)\,
\frac{H^{(2)}_{\ell}(ka)}{H^{(1)}_{\ell}(ka)} .
\label{eq:SnmI}
\end{eqnarray}
Taking into account that $ |S^{\rm (O)}_{nm}| = 1 $, we have $ |S_{nm}| = |S^{\rm (I)}_{nm}| $, and as shown in Fig.~\ref{fig:Frischat_Doron}, $ S $ becomes almost diagonal. 
The region inside the dashed line indicates the classically allowed region, while the classically forbidden region extends outside it. 
For fixed $ n $, one can see that $ |S_{nm}| $ decays faster than exponentially as a function of $ m $.

By applying the Poisson summation formula to the expression for $S^{\rm (I)}_{nm}$ in Eq.~(\ref{eq:SnmI}), we obtain  
\begin{eqnarray}
S^{\rm (I)}_{nm}
=
-\sum_{\mu=-\infty}^{\infty}\int_{-\infty}^{\infty}
d\ell\;
e^{\,i2\pi\mu\ell + i\pi(m-\ell)}\,
J_{n-\ell}(k\delta)\,J_{\ell-m}(k\delta)\,
\frac{H^{(2)}_{\ell}(ka)}{H^{(1)}_{\ell}(ka)}. 
\label{3}
\end{eqnarray}
For the case $ \mu = 0 $, one can rewrite the expression by using Sommerfeld's integral representation for the Bessel functions and the Debye approximation for the Hankel functions, leading to an integral representation of $S^{\rm (I)}_{nm}$.  
Furthermore, by applying the saddle-point approximation to the integral representation thus obtained under the condition $k \gg 1$, one arrives at the following semiclassical expression for the $S$-matrix:
\begin{eqnarray}
S^{\rm (I)}_{nm} \approx
\sum_{p}
\sqrt{\frac{|\mathcal{R}_p|}{k}}\;
\exp\!\left[ i k \Phi_p + \frac{i}{2}\arg \mathcal{R}_p - \frac{3\pi i}{4} \right], 
\label{eq:semiapproxS}
\end{eqnarray}
where the phase $\Phi_p$ represents the reduced action of the ray and $|\mathcal{R}_p|$ is the corresponding reciprocal stability.  Therefore, the formula (\ref{eq:semiapproxS}), which is given in terms of the sum over classical orbits, can be viewed as an analog of Van-Vleck formula in the standard semiclassical formulation~\cite{van1928correspondence}.

By performing a saddle-point approximation, 
trajectories inside the billiard can be analytically continued into the complex coordinates, thereby allowing the evaluation of tunneling contributions.
To describe the interaction with the inner circle, we consider the complex circle of radius $a$ centered at the point $(x_0, y_0)$:
$x - x_0 = a\cos\beta,\ \ y - y_0 = a\sin\beta$. 
When $\beta$ is real, the corresponding point lies on the inner circle and yields trajectories that reflect at the billiard boundary. 
In contrast, by allowing $\beta$ to take complex values, the collision point on the inner circle can be analytically continued into the complex plane, and the resulting trajectories propagate in the complex domain. 
Note that extending $\beta$ into the complex plane is exactly the same operation as extending the initial phase $\theta$ to a complex value when performing the time-domain semiclassical analysis closely discussed in Subsec.~\ref{sec:Time-domain}.
The ``classical trajectories'' contributing to the sum in (\ref{eq:semiapproxS}) are those that start with the initial collision parameter $L_i = n/k$, are reflected at the complex inner circle, and end with the final collision parameter $ L_f = m/k$. 
In other words, the saddle-point condition of the integral is equivalent to the reflection condition at the scatterer surface, analytically continued into the complex domain.

According to this recipe, one finds multiple saddles in the complex plane in general. 
However, not all complex trajectories that satisfy the saddle-point condition finally contribute to the sum in (\ref{eq:semiapproxS}). 
There are two reasons for this. 
First, although the reduced action $\Phi_p$ of the contributing complex trajectories has an imaginary part, those for which the imaginary part is large and positive give contributions that are exponentially smaller in the sum than those with a smaller imaginary part. Conversely, trajectories whose imaginary part is large and negative would yield exponentially large contributions, but such complex trajectories should not appear in the saddle-point sum because they correspond to unphysical contributions. 
More precisely, trajectories for which the imaginary part becomes large and negative are eliminated by the Stokes phenomenon, but we will not go into this issue here. For further details, see Refs.~\cite{dingle1973,ecalle1981,voros1983,berry1988stokes,olver1997asymptotics,delabaere1997,bender1999advanced,balser2006divergent,mitschi2016divergent}.

In \cite{doron1995semiclassical,frischat1998dynamical}, the authors demonstrated that a single dominant saddle, depicted in Fig.~\ref{fig:Frischat_Doron} as gray branches emanating from the real contributions, reproduces the exact quantum result reasonably well.
The treatment that includes the dominant tunneling branch attached to the real manifold is exactly the same as that used in the {\it one-step} time domain semiclassical calculation~\cite{shudo1995,shudo1998}. 
A similar treatment was also employed in \cite{mertig2013complex}. 

One of the major advantages of the scattering matrix approach is that the tunneling splitting can be approximately expressed in terms of the scattering matrix $S_{nm}$. 
After assessing the validity of several approximations, one can reach the following expression:
\begin{eqnarray}
\delta\theta_n \approx \frac{2}{N} \Bigl|[S^N]_{-n,n} \Bigr| = \frac{2}{N} \left| \sum_{\{\lambda_i\}} \prod_{i=1}^{N-1} S_{\lambda_i,\lambda_{i+1}} \right|.
\end{eqnarray}
Here, the sum over intermediate states $\{\lambda_i\}$ runs over all paths $\{\lambda_i\}_{i=1}^{N}$ in the space of matrix elements from $\lambda_1 = -n$ to $\lambda_N = n$. 
Moreover, to obtain reasonable results for tunneling splitting, one must take $N$ sufficiently large \cite{hackenbroich1998quantum}.

The simplest among such paths is the one that involves only the direct tunneling amplitude $ S_{-n,n} $. Specifically, the path satisfying the conditions, 
\begin{eqnarray*}
  \lambda_1=\lambda_2=\cdots=\lambda_i=-n,\qquad
  \lambda_{i+1}=\lambda_{i+2}=\cdots=\lambda_{N-1}=n, 
\end{eqnarray*}
represents a direct tunneling path. 
Note that this mechanism is very close to Wilkinson's scenario~\cite{takada1995transfer}. 
The magnitude of such a direct path contribution is, however, extremely small; a much larger contribution comes from paths that tunnel only a short distance in angular-momentum space and traverse the remaining region via classically allowed transitions. 

As is clear from the semiclassical analysis, the closer the target state $m$ is to the regular state $n$, that is, the smaller $|m-n|$ is, the weaker the exponential attenuation. 
In contrast, transitions within the chaotic region do not suffer exponential decay since the transition proceeds on the real plane. 
Consequently, tunneling paths that jump to states as close as possible to the regular state tend to dominate. 
This precisely matches the scenario envisaged by chaos-assisted tunneling explained in Subsec.~\ref{sec:CAT}. 
The direct path represents the tunneling trajectory in the absence of chaos, whereas the path going through the chaotic region represents the trajectory associated with chaos-assisted tunneling.

Furthermore, in Refs.~\cite{doron1995semiclassical,frischat1998dynamical} the authors carried out an analysis of tunneling paths that takes into account the structure of the chaotic region.
As is well known, the orbits in chaotic seas tend to stick near regular regions, and partial barriers nearly partition the phase space, resulting in non-uniform transport within the chaotic region.
Under such circumstances, one can expect two types of chaos-assisted paths to make the dominant contribution to the tunneling splitting.

\bn
(I) $ |\, n\, \rangle \to |\, \gamma\, \rangle \to$ $|\, $-$n\, \rangle$ : a path in which the torus state $|\,n\,\rangle$ tunnels directly to a chaotic state $ |\,\gamma\,\rangle $, evolves within the chaotic region, and then tunnels to the opposite torus state $|\,$-$n\,\rangle$.

\n
(II) $ |\, n\, \rangle \to |\, l\, \rangle \to |\, \gamma\, \rangle \to |\,$-$l\, \rangle \to$$|\, $-$n\, \rangle$: a path in which the transition from the torus state $ |\, n\, \rangle $ to the opposite torus state |\,-$n\, \rangle$ proceeds via the edge states $ |\, l\, \rangle $ and $ |\,$-$l\, \rangle $ located between the torus and chaotic states, with an intermediate visit to a chaotic state $ |\, \gamma\, \rangle $.

\b

Comparing the magnitudes of these two contributions, one finds that
$ \delta\theta_n^{(\mathrm{II})} \gg \delta\theta_n^{(\mathrm{I})} $. 
This follows from the fact that, for edge states, one typically has $ |\, n-l\, | > |\, n-\gamma\, | $, so the exponential decay of the former is weaker than that of the latter. 
Hence, they emphasized the importance of tunneling paths that proceed via such edge states—dubbed {\it beach states} in their papers~\cite{doron1995semiclassical,frischat1998dynamical}. 
This observation is consistent with the one made based on the fully semiclassical analysis using complex trajectories, which will be discussed in Subsec.~\ref{sec:Most_dominant}. 

The above approach evaluates the one-step time evolution using the saddle-point approximation, which 
yields the formula (\ref{eq:semiapproxS}), and then examines the energy splitting by propagating the resulting $S$-matrix quantum mechanically. 
From a semiclassical perspective, this may be regarded as a hybrid approach,
with one part of the analysis carried out classically and the other quantum mechanically.
Thus, this does not amount to an understanding of tunneling based entirely on classical dynamics. 
An analysis employing a fully semiclassical treatment in the time domain will be introduced in detail in Sec.~\ref{sec:Complex_path}.

\section{Role of classical and quantum resonances}
\label{sec:Role_resonance}

\subsection{Avoided crossings and classical resonances}
\label{sec:Avoided_crossings}

Under a small perturbation of an integrable system, the Poincar\'e-Birkhoff theorem predicts that each resonant torus breaks up into an alternating chain of stable and unstable periodic points.
In mixed phase space, as illustrated in Fig.~\ref{fig:map_splitting}(a), not only chaotic regions but also the nonlinear resonances formed in this way are characteristic invariant structures, and their influence on dynamical tunneling is of interest.

We begin with a heuristic argument illustrating how classical nonlinear resonances manifest themselves in the corresponding quantum system.
To this end, consider a two-dimensional Hamiltonian expressed in terms of pairs of action-angle variables:
\begin{eqnarray}
H(I_1, I_2,\theta_1, \theta_2) = H_0(I_1, I_2) + \varepsilon H_1(I_1, I_2,\theta_1, \theta_2), 
\end{eqnarray}
and suppose that the angular frequencies of the completely integrable part $H_0(I_1, I_2)$
\begin{eqnarray}
\omega_i(I_1, I_2) = \frac{\partial H_0(I_1, I_2) }{\partial I_i}, ~~~~~~(i = 1, 2) 
\end{eqnarray}
depend nonlinearly on the action variables $(I_1, I_2)$. 
The ratio $\omega_1(I_1, I_2)/\omega_2(I_1, I_2)$ in this case varies continuously (analytically in most cases) as a function of $(I_1, I_2)$. 
If the perturbation strength $\varepsilon$ is sufficiently small and the ratio $ \omega_1(I_1,I_2)/\omega_2(I_1,I_2) $ is an irrational number sufficiently far from rationals (more precisely, it should be Diophantine), the quasi-periodic motions of the integrable Hamiltonian $ H_0(I_1,I_2) $ persist, according to the Kolmogorov-Arnold-Moser (KAM) theorem.
In contrast, if the ratio $\omega_1(I_1,I_2)/\omega_2(I_1,I_2)$ is rational, such orbits are fragile under perturbations; for generic perturbations, the Poincar\'e-Birkhoff theorem leads to the emergence of a pair of stable (elliptic) and unstable (hyperbolic) periodic orbits~\cite{lichtenberg2013regular,de1988hamiltonian}.
In general, resonance phenomena exhibited by nonlinear systems are called \textit{nonlinear resonances}. In what follows, we use the term ``nonlinear resonance'' specifically to refer to the island structures that appear around stable (elliptic) periodic orbits in phase space.

In this Section, we investigate how classical nonlinear resonances influence quantum tunneling. 
The ways in which nonlinear resonances and chaos manifest in the corresponding quantum dynamics have long been studied, particularly in the field of chemical physics~\cite{keshavamurthy2005dynamical,keshavamurthy2011dynamical,keshavamurthy2007dynamical}. 
As the classical dynamics undergoes a transition from regular to chaotic behavior, both the character and the frequency of avoided crossings in the energy spectrum change; these changes are recognized as quantum manifestations of classical non-integrability~\cite{noid1980properties,noid1980calculations,ramaswamy1981onset}.
The overlap, or accumulation, of avoided crossings in the energy spectrum has been proposed as a fingerprint of the breakdown of classical invariant tori, and hence of classical chaos.
On the other hand, within a primitive semiclassical framework, Markus and co-workers investigated the origin of avoided crossings, and associated avoided crossings with classically forbidden processes, not merely over-the-barrier tunneling, but rather with dynamical tunneling~\cite{ramaswamy1981perturbative,noid1983comparison,uzer1983uniform,ozorio1984tunneling,heller1995dynamic}.

The association between classical nonlinear resonances 
and avoided crossings can be made in the following way~\cite{ramachandran1993influence}. 
Let us consider, for simplicity, a two-degree-of-freedom system with Hamiltonian $H(I_1,I_2,\lambda_0)$ that depends on two action variables $I_1, I_2$ and a parameter $\lambda$. 
If quantum levels degenerate at $\lambda = \lambda_0$, 
the situation can  be expressed semiclassically as $H(I_1,I_2,\lambda_0) = H(I_1',I_2',\lambda_0)$. 
Provided that $|I_n - I'_n| \ll1 ~~(n=1,2)$, we can expand $H(I_1',I_2',\lambda_0)$ 
around $I'_n = I_n$ to obtain
\begin{align}
\nonumber
 \label{eq:expansion}
 H(I_1',I_2',\lambda_0) &= H(I_1,I_2,\lambda_0)  
+ (I_1-I'_1) \omega_1 + (I_2-I'_2) \omega_2 + \cdots, 
\end{align}
where $\omega_n = \partial H/\partial I_n~(n=1,2)$. 
Now assuming the semiclassical quantization condition $I_n =( m + \alpha_n/4) \hbar$
where $\alpha_n$ is the Maslov index, 
we find the condition at $\lambda = \lambda_0$,
\begin{align}
r \omega_1= s\omega_2, 
\end{align}
where $r =m_1 - m'_1$ and $s = m_2 - m'_2$, 
which is exactly the classical resonance condition. 
Therefore, degeneracy or interaction between two quantum states implies the presence of a resonance in the corresponding classical system.
Such interaction generally lifts the degeneracy, giving rise to a small level splitting and hence an avoided crossing.
Thus, the occurrence of avoided crossings may be linked to classical nonlinear resonances.

However, the above argument suggests only a possible correspondence between avoided crossings and nonlinear resonances. Not every avoided crossing necessarily corresponds to a specific nonlinear resonance~\cite{ramachandran1993influence}, and it should be noted that the correspondence between classical nonlinear resonances and avoided crossings remains under investigation~\cite{wisniacki2011poincare,arranz2021correspondence}.
Nevertheless, given the possibility that avoided crossings manifest tunneling, particularly dynamical tunneling~\cite{noid1983comparison,uzer1983uniform,ramaswamy1981perturbative,heller1995dynamic,ozorio1984tunneling}, and in light of the correspondence between classical nonlinear resonances and avoided crossings, it is natural to expect that classical nonlinear resonances affect dynamical tunneling.

\subsection{Recipe for resonance-assisted tunneling}
\label{sec:RAT}

As explained in Sec.~\ref{sec:Role_chaos}, CAT can manifest itself as resonant spikes due to avoided crossings between a tunneling doublet and a third state supported by the chaotic region.
On the other hand, Bonci et al. discovered that the states interacting with a tunneling doublet can, in some cases, be localized on nonlinear resonances in phase space~\cite{bonci1998tunneling}. 
This connects with earlier work in chemical physics and suggests that classical nonlinear resonances may also influence dynamical tunneling.
One should also keep in mind that spikes in the tunneling splitting are not restricted to non-integrable systems.
Le Deunff et al. pointed out that, in a one-dimensional symmetric triple-well system, sharp spikes similarly appear in the energy splitting as system parameters are varied, owing to the interaction between a tunneling doublet localized in the two outer wells and a state localized in the central well~\cite{le2010instantons}.
The appearance of spikes in integrable systems is seen similarly in other systems~\cite{le2013,hanada2015}

RAT theory provides a procedure for evaluating tunneling rates or tunneling splittings by applying quantum perturbation theory to a local Hamiltonian constructed from classical nonlinear resonances.
RAT is a hybrid approach in the sense that it is a quantum perturbation theory employing information from classical phase space. 
The system under consideration in the RAT argument is a one-dimensional system 
subject to an external periodic drive:
\begin{eqnarray}
H(I,\theta,t) = H_0(I) + V(I,\theta,t). 
\label{eq:}
\end{eqnarray}
Although a theory incorporating classical nonlinear resonances should not be, in principle, restricted to periodically driven systems, 
as will be discussed in detail in Subsec.~\ref{sec:QR}, it turns out that the periodic driving force plays a crucial role in understanding the observed tunneling phenomena.
For the moment, we set this point aside and present the prescription proposed in Refs.~\cite{brodier2001,brodier2002}.

We consider a Hamiltonian in the action-angle form since 
we here focus on an island structure associated with a prominent, say $r\!:\!s$, resonance in phase space, and assume that the motion in its vicinity is approximated via the standard secular perturbation theory. 
The $r:s$ resonance occurs when $s$ internal oscillation periods match $r$ driving periods, and 
$r$ sub-islands appear in the stroboscopic phase space. 
Here, the system is assumed to consist of an integrable Hamiltonian $H_0$ which provides an approximation of the regular motion and a weak perturbation $V$. 

Suppose that the nonlinear $r\!:\!s$ resonance appears at the action variable $I_{r:s}$ that satisfies the condition,
\begin{eqnarray}
r \, \Omega_{r:s} = s \omega,
\label{eq:rs-condition}
\end{eqnarray}
with $\omega = 2\pi/\tau$ and
\begin{equation}
\Omega_{r:s} \equiv \left.\frac{dH_0}{dI}\right|_{I = I_{r:s}} .
\end{equation}

By introducing a co-rotating frame moving with the resonance,
$\vartheta := \theta - \Omega_{r:s}\, t$,
the variable $\vartheta$ becomes slow. This allows us to apply an adiabatic approximation, and the system can then be reduced to a one-dimensional time-independent Hamiltonian.
Expanding the part $ H_0(I) $ to the lowest order around $ I = I_{r:s} $, one finally finds the Hamiltonian near the $ r\!:\!s $ resonance,
\begin{eqnarray}
H_{\text{res}}(I,\vartheta) \simeq \frac{(I - I_{r:s})^2}{2 m_{r:s}} + \sum_{k=1}^{\infty} 2 V_k(I)\cos(kr\,\vartheta + \phi_k).
\label{eq:Hres}
\end{eqnarray}
Here, $ V_k $ is defined from the Fourier expansion of $ V(I,\theta,t) $,
\begin{eqnarray}
V(I,\theta,t) = \sum_{l,m=-\infty}^{\infty} V_{l,m}(I) \, e^{i l \theta} e^{i m \omega t}\. , 
\label{eq:V-Fourier}
\end{eqnarray}
and 
\begin{eqnarray}
V_k(I) := V_{r k,\,-s k}(I)\, e^{-i\phi_k}, 
\label{eq:Vk-definition}
\end{eqnarray}
where $ \phi_k $ denotes the phase that appears in the Fourier expansion of $ V(I,\vartheta,t) $ after averaging over the interval $ r\hspace{0.2mm}\tau $.

If the potential term $ V $ is small enough, we may apply quantum perturbation theory to the Hamiltonian (\ref{eq:Hres}).
The eigenstates $|\,\psi_n\rangle$ of $\hat{H}_{\text{res}}$ 
can be expanded in terms of unperturbed states $|\,n'\,\rangle$ that satisfy the selection rule $|\,n' - n\,| = kr$ with integer $k$: 
\begin{eqnarray}
|\,\psi_n\rangle &=& |\,n\rangle + \sum_k \frac{\langle \,n+kr\, | \,\hat{H}_{\text{res}}\, |\, n \,\rangle}{E_n - E_{n+kr} + ks \hbar \omega} |\,n+kr\,\rangle \nonumber \\
&+& \sum_{k,k'} \frac{\langle \,n+kr \,|\, \hat{H}_{\text{res}} \,|\, n + k'r \,\rangle}{E_n - E_{n+kr} + ks \hbar \omega} \cdot
\frac{\langle \,n+k'r\, |\, \hat{H}_{\text{res}} \,|\, n \,\rangle}{E_n - E_{n+k'r} + k's \hbar \omega} |\,n+kr\,\rangle + \cdots ,
\label{eq:psi-expansion}
\end{eqnarray}
Note that if the potential $ V $ is analytic, the coefficients in its Fourier expansion (\ref{eq:Vk-definition}) decay exponentially. Consequently, we can expect that the direct first-order perturbative term connecting the states $ |\,n\,\rangle $ and $ |\,n+kr\,\rangle $, expressed as a single sum over $k$, can be smaller than the second- and higher-order perturbative contributions. 
In addition, for quantitatively accurate calculations, one must also take into account the action dependence of $ V_k(I) $~\cite{lock2010,Schlagheck11}.

Actual calculations are performed using the simpler Hamiltonian,
\begin{eqnarray}
H_{\text{res}}(I,\vartheta) \simeq \frac{(I - I_{r:s})^2}{2 m_{r:s}} + 2 V_{r:s} \cos(r \vartheta + \phi_1). 
\label{eq:Hres-effective}
\end{eqnarray}
Here, the parameters $ I_{r:s} $, $ m_{r:s} $, and $ V_{r:s} $ are determined numerically from the classical phase space generated by the corresponding classical map. Note that this is precisely the moment at which information from the classical phase space is incorporated. 
As is now clear, the RAT prescription constructs a local one-dimensional Hamiltonian around a targeted nonlinear resonance using classical phase-space data, and then applies quantum perturbation theory to the resulting local Hamiltonian. 
The term ``hybrid'' classical--quantum method refers to this combination.

We make an important observation for the case where it suffices to consider only the first-order perturbative term.
Within the quadratic approximation of $H_0(I)$ around $I_{r:s}$ the energy differences are expressed as 
\begin{equation}
E_n - E_{n+kr} + ks \hbar \omega \simeq \frac{1}{2m_{r:s}} (I_n - I_{n+kr})(I_n + I_{n+kr} - 2 I_{r:s}).
\label{eq:energy-difference}
\end{equation}
From this, we see that the coupling between $|\,n\,\rangle$ and $|\,n'\,\rangle$ becomes particularly strong if the $r\!:\!s$ resonance is symmetrically located between the two tori that are associated with the actions $I_n$ and $I_{n'}$. 
Such a situation occurs when the condition $I_n + I_{n'} \simeq 2 I_{r:s}$ is satisfied. 
Alternatively stated, the RAT mechanism assumes that the local ``ground state", which is localized in the center of the regular region (with action variable $I_0 < I_{r:s}$) is coupled to a highly excited state (with action variable $I_{kr} > I_{r:s}$) via the $r\!:\!s$ nonlinear resonance. 
In this manner, the two states mediated by the nonlinear resonance are coupled, and this represents 
the core mechanism of RAT.
A fully semiclassical, as opposed to hybrid, treatment representing this situation was presented in the appendix of Ref.~\cite{brodier2002}. 
Because the local Hamiltonian~(\ref{eq:Hres-effective}) is one-dimensional, the complex trajectory linking the two classically disconnected invariant curves mediating the nonlinear resonance is essentially the instanton path. 
A similar situation is observed in a semi-global setting realized by a one-dimensional normal form Hamiltonian~\cite{le2013}. 
We will discuss this point in detail below.

\subsection{Prediction based on the RAT approach}
\label{sec:Verification_of_RAT}

\begin{figure}[ht]
\centering
\includegraphics[width=0.80\linewidth, trim=0 105mm 0 110mm, clip]{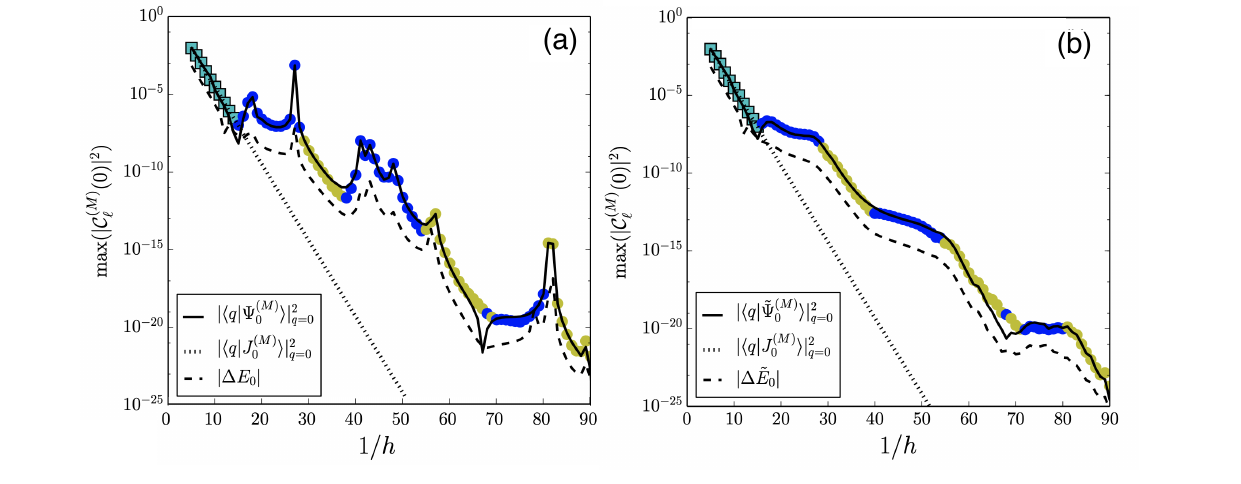}
\caption{
	\label{fig:maximal_mode} 
	(a) Tunneling splittings (dashed curve) and maximal modes (squares) plotted as a function of the inverse Planck's constant $1/\hbar$.
	Cyan squares show the instanton regime~\cite{hanada2015,hanada2023dynamical}.
	Blue and yellow dots indicate the regimes in which the maximal-mode energy lies above and below the separatrix energy, respectively.
	(b) Maximal modes for the absorbed eigenstates $\ket{\,\tilde \Psi_n\,}$.
	The absorption procedure is explained in the text. 
	The absorption parameters are chosen as $s=3$ and $\Gamma=0.4$.
	The rule for color coding is the same as in (a).
	In both calculations, the BCH order $M=7$ was used.
	In (a) the exact eigenfunction $\bracket{\,q\,}{\,\Psi_0^+\,}$ at $q=0$, 
	integrable basis  $\bracket{\,q\,}{\,J_n^{(M)}\,}$, and energy splitting $\Delta E_0$ are shown as solid,
	dotted and dashed curves, respectively. 
	In (b) the solid curve represents the absorbed eigenfunction 
	$\bracket{\,q\,}{\,\tilde \Psi_0^+\,}$ at $q=0$, 
	and dotted and dashed ones are the same as in (a).
	Adapted from Ref.~\cite{hanada2015}. 
}
\end{figure}

In this Subsection, we examine what insights the RAT approach can provide into the nature of tunneling. 
We discuss this issue again by employing the quantum map~(\ref{eq:qmap})
and the quasi-eigenenergies of the unitary operator $\hat U$. 
Here, the energy splitting is defined as $\Delta E_n := E_n^{+} - E_n^{-}$, where $ E_n^{+} $ and $ E_n^{-} $ denote the quasi-eigenenergies corresponding to even and odd parity, respectively.
To see how the tunneling probability is enhanced,
it is common to plot the energy splitting or the tunneling rate as a function of $1/\hbar$~\cite{roncaglia1994,brodier2001,brodier2002,Mouchet03,mouchet2006influence,mouchet2007importance,backer2008regular,backer2010direct,lock2010,mertig2013complex,hanada2015,hanada2023dynamical}.
Before turning to the RAT approach, we note that the tunneling splitting exhibits a characteristic step-like structure as a function of $1/\hbar$, as illustrated in Fig.~\ref{fig:maximal_mode}.

The plot of Fig.~\ref{fig:maximal_mode} is obtained as follows. 
First, recall that the tunneling energy splitting approximately reflects the value of the wavefunction $|\Psi_0\rangle$ at $q =0 $~\cite{hanada2015}.
To explore the nature of wavefunctions at $q=0$, 
we introduce a spectrum decomposition at each position $q$ 
in terms of integrable bases $\ket{\,J^{(M)}_\ell}$: 
\begin{equation} \label{eq:coeff_expand}
  \bracket{\,q\,}{\,\Psi^{+}_0\,} = \sum_{\ell=0}^{N-1}  \mathcal{C}_{\ell}^{(M)}(q),
\end{equation}
where
\begin{equation} \label{eq:cont}
 \mathcal{C}_{\ell}^{(M)}(q) := \bracket{\,q\,}{\,J^{(M)\, }_\ell} \bracket{\,J^{(M)\,}_\ell}{\,\Psi_0^{+}\,}.
\end{equation}
Such a decomposition was called the {\it contribution spectrum} in Refs.~\cite{shudo2014instanton,hanada2015}.
Here, $ \ket{\,J^{(M)}_\ell} $ denotes an eigenstate of the integrable Hamiltonian $ \hat{H}_\mathrm{eff}^{(M)} $:  
\begin{equation}
\label{eq:BCH_eigenstate}
  \hat{H}_\mathrm{eff}^{(M)}\ket{\,J^{(M)}_\ell}   =  E_{\ell}^{(M)} \ket{\,J^{(M)}_\ell}.
\end{equation}
The integrable Hamiltonian $ \hat{H}_\mathrm{eff}^{(M)} $ is obtained by approximating the time-evolution operator $ \hat{U} $ of the system using the Baker-Campbell-Hausdorff (BCH) expansion as  
\begin{equation}
  \hat{U}  \approx 
  \hat{U}_M := \exp\Bigl[-\frac{i}{\hbar}\tau \hat{H}_\mathrm{eff}^{(M)}(\hat{q},\hat{p})\Bigr],
\end{equation}
and the Hamiltonian $\hat{H}_\mathrm{eff}^{(M)}(\hat{q},\hat{p})$ obtained by truncating the BCH series is expressed explicitly as
\begin{equation}\label{eq:bch_hamiltonian}
  \hat{H}_\mathrm{eff}^{(M)}(\hat{q},\hat{p})
  =\hat{H}_1(\hat{q},\hat{p})+\sum_{\underset{(j\in \text{odd int.})}{j=3}}^{M}
  \biggl(\frac{i\tau}{\hbar}\biggr)^{j-1}\hat{H}_{j}(\hat{q},\hat{p}).
\end{equation}
Here $\hat{H}_j$ denotes the $j$-th order term in the BCH series.

From Fig.~\ref{fig:maximal_mode}, we notice that the value of the eigenstate $\ket{\,\Psi(q)\,}$ at $q=0$,
and hence the behavior of the energy splitting, is well approximated by the maximal mode,
namely $\displaystyle \max_{\ell}\, \bigl|\,\mathcal{C}_{\ell}^{(M)}(0)\,\bigr|^2$.
Note that the maximal mode, as discussed in detail later, corresponds to a {\it quantum-resonant state}.
It is also important to note the $1/\hbar$ dependence of the location of this maximal mode.
In Fig.~\ref{fig:maximal_mode}, blue circles indicate that the maximal mode lies {\it outside the separatrix},
whereas yellow markers indicate that it lies {\it inside} the separatrix.
Accordingly, the repeating sequence ``plateau $\to$ exponential decay $\to$ plateau $\to \cdots$'' observed in Fig.~\ref{fig:maximal_mode} 
is explained by the maximal mode successively moving as ``outside $\to$ inside $\to$ outside $\to \cdots$''.

The fact that, in the plateau region, the maximal mode located outside determines the value of the wavefunction at $q=0$ can also be confirmed by focusing on the behavior of the wavefunction at $q=0$~ \cite{hanada2015,hanada2023dynamical}. 
In the plateau region, the wavefunction at $q=0$ becomes flat rather than valley-shaped; this reflects the fact that the outer (rotational) component dominates the behavior of the wavefunction at $q=0$ (see Fig.~\ref{fig:hump}). 
These observations indicate that the plateau regions appear because the ground state couples strongly 
to outer rotational states.

\begin{figure}[ht]
\centering
\includegraphics[width=0.80\linewidth, trim=0 100mm 0 110mm, clip]{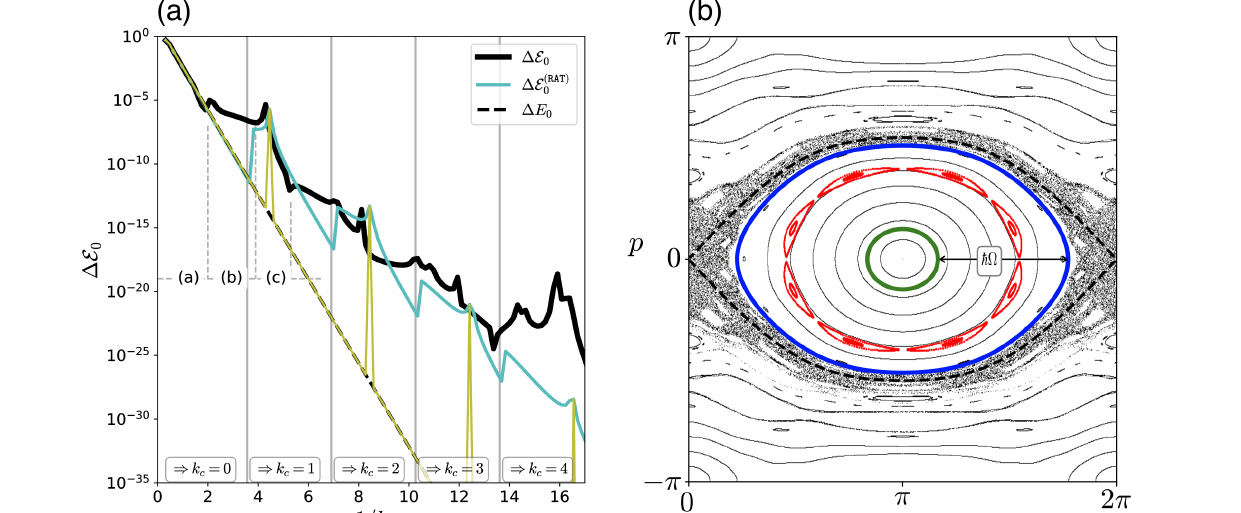}
\caption{
\label{fig:RAT_test}
  	(a) The black solid and dashed curves indicate tunneling splitting $\Delta \E_0$ obtained by the exact calculation 
	and $\Delta E_0$ obtained by diagonalizing the BCH Hamiltonian.   
	The cyan curve is the tunneling splitting 
	$\Delta \E_0^{(\mathtt{RAT})}$ obtained based on the RAT recipe. 
	The gray vertical lines show the values of $1/\hbar$ at which the value of $k_c$ given by 
	Eq.~(\ref{eq:kc}) is incremented by one.
	For further details, see Ref.~\cite{hanada2023dynamical}. 
	(b) Phase space portraits of the classical map $f$ with $\eps=(15/16)^2$. The classical resonance 
    with $r:s=8:1$ is shown as red curves. 
    The energy contours of the BCH Hamiltonian $H_\textrm{cl}$ associated with the RAT scheme are drawn as solid curves (see the legend in the figure).
    The separatrix of the BCH Hamiltonian is shown by the black dashed curve.
    Adapted from Ref.~\cite{hanada2023dynamical}. 
}
\end{figure}

With this in mind, we now examine in more detail what happens when we carry out the RAT calculation.
As illustrated in Fig.~\ref{fig:RAT_test}, we observe a monotonic exponential decay in the small $1/\hbar$ region ($0 \lesssim 1/\hbar \lesssim 1.6$). This region is often referred to as the \textit{direct-tunneling} or \textit{instanton} regime, where an integrable approximation works and successfully yields the tunneling rate~\cite{backer2008regular,backer2008dynamical,backer2010direct,shudo2014instanton}.
In this sense, no clear signatures of non-integrability can be found there. 
As $1/\hbar$ increases, a plateau emerges in the $\Delta E_n $ vs. $ 1/\hbar$ plot ($1.6 \lesssim 1/\hbar \lesssim 4.5$). 
Further increasing $1/\hbar$ reveals another region where the energy splitting again decreases exponentially ($4.5 \lesssim 1/\hbar \lesssim 5.1$).  
Hereafter we refer to these three regions as (a) the \textit{first decay}, (b) the \textit{plateau}, and (c) the \textit{second decay region}, respectively.

Next, we examine what the RAT recipe predicts for the plateau and the second decay region. 
To this end, we employ the formula (\ref{eq:psi-expansion}), further refined to take into account the action dependence of $V(I)$~\cite{Schlagheck11}:
\begin{equation}\label{eq:rat_wave}
\ket{\Psi_n^{\mathtt{(RAT)}}} = \ket{J_n} + \sum_{k>0} B_{n+kr,n} \ket{J_{n+kr}},
\end{equation}
where
\begin{equation}
\label{eq:B_coeff}
B_{n+kr,n} = \prod_{\ell=1}^{k_c}  \frac{A_{n+\ell r,\,n+(\ell-1)r}}{E_n - E_{n+\ell r} + \ell s\,\hbar \omega},
\end{equation}
and
\begin{equation}\label{eq:rat_coeff}
A_{n+\ell r,\,n+(\ell-1)r} = V_{r:s}(I_{r:s})\, e^{i\phi_k} \biggl(\frac{\hbar}{I_{r:s}}\biggr)^{kr} \sqrt{\frac{(n+kr)!}{n!}}.
\end{equation}
Here, $k_c$ is determined by the formula, 
\begin{equation}
\label{eq:kc}
k_c = \left\lfloor\frac{1}{r}\biggl( \frac{\mathcal{A_\mathrm{reg}}}{2\pi\hbar} - \frac{1}{2} \biggr) \right\rfloor.
\end{equation}
$\mathcal{A}_\mathrm{reg}$ denotes the area occupied by the regular region.

First, it is evident that this expression cannot account for the energy splitting in the plateau region. 
The reason is that the sum in the RAT formula (\ref{eq:B_coeff}) has an upper cutoff, 
but we have $k_c = 0$ in the plateau region. 
The case $k_c = 1$ is reached only at the right edge of the plateau; that is, at the left edge of the second decay region. 
As indicated above, the plateau corresponds to the region in which the maximal mode lies outside the separatrix.
Since the RAT approach is formulated in terms of a pendulum-type Hamiltonian~(\ref{eq:Hres-effective}),
couplings to outer (rotational) states fall outside its scope.
Accordingly, RAT calculations do not reproduce the plateau. 

On the other hand, as seen in Fig.~\ref{fig:RAT_test}, a spike appears at the moment when the plateau switches to the second decay region. 
As discussed in Subsec.~\ref{sec:RAT}, spikes arise when a third state approaches the tunneling doublet under consideration and forms an avoided crossing. 
In the RAT calculation, spikes appear when the condition $I_n + I_{n'} \simeq 2 I_{r:s}$ is satisfied as mentioned in the end of  Subsec.~\ref{sec:RAT}. In that case, the energy denominator becomes nearly zero and a resonance occurs.

Figure~\ref{fig:RAT_test}(b) shows the corresponding classical phase space.
The RAT scheme first requires finding visible nonlinear resonances in the region enclosed by the separatrix. 
In the present case, the $r:s = 8\!:\!1$ resonance chain is the lowest resonant condition and most visible. 
Of course, there should be infinitely many nonlinear resonances buried in the regular region. 
However, if we require that the size of nonlinear resonances should be comparable to the size of the Planck cell, then the $r:s = 8\!:\!1$ resonance is the only candidate expected to allow RAT contributions. 

As seen in Fig.~\ref{fig:RAT_test}, the RAT calculation reproduces well the spikes that appear when the plateau switches to the second decay. 
Moreover, if one simply continues to include the perturbative term obtained for $k_c = 1$, the resulting curve reasonably reproduces the exact energy splittings for the second  decay region. 
However, it should be noted that, away from the spike, the condition $ I_n + I_{n'} \simeq 2 I_{r:s} $ is no longer fulfilled. 
RAT provides a tunneling coupling, mediated by classical nonlinear resonance, between states of equal action located on opposite sides of the resonance.
This is the ``classical'' mechanism envisioned by RAT, and it is precisely what the condition $I_n + I_{n'} \simeq 2 I_{r:s}$ implies. 
As long as one works within the pendulum Hamiltonian framework (\ref{eq:Hres-effective}), a tunneling coupling linked to the classical nonlinear resonance cannot arise between states that do not satisfy this  condition.

\begin{figure}[ht]
\centering
\includegraphics[width=0.80\linewidth, trim=0 100mm 0 100mm, clip]{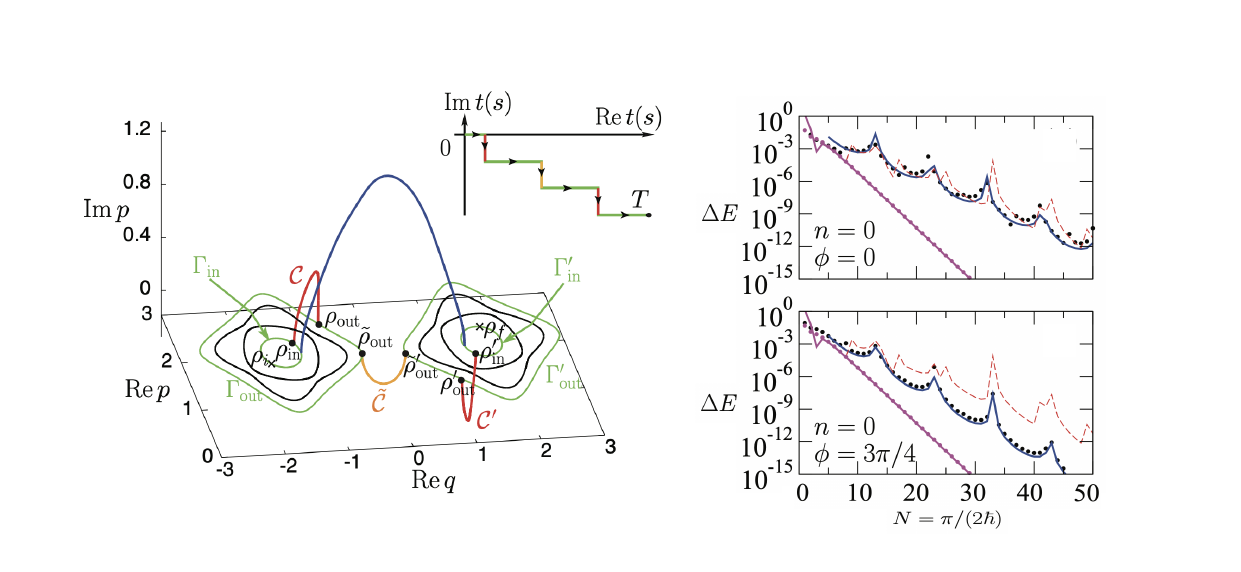}
\caption{
\label{fig:Mouchet}
Left panel: 
Illustration of the real phase space (black and green) and the complex paths (blue, orange, and red)
connecting two disjoint energy surfaces, $\Gamma_{\rm in}$ and $\Gamma'_{\rm in}$.
Two types of complex paths appear: a dark-blue (direct) path and a composite path
$C$ (red) $+\ \tilde{C}$ (orange) $+\ C'$ (red).
The latter constitutes a complex trajectory generated by the time contour depicted in the upper-right inset.
Right panel: Quantum and semiclassical level splittings plotted in a semilogarithmic scale, as a function of the integer $N = \pi / 2\hbar$. The (black) dots represent the exact numerical results while the (blue) solid lines show the predictions obtained by the semiclassical formula.
The (red) dashed line is the perturbative RAT prediction, which fails to reproduce the quantum result in the lower panel case. 
Ref.~\cite{le2013} developed an argument for why the RAT calculation does not work.
Reproduced from Ref.~\cite{le2013}. 
}
\end{figure}


In fact, the normal-form Hamiltonian studied in Ref.~\cite{le2013semiclassical} is precisely intended to model such a situation.
In Ref.~\cite{le2013semiclassical}, tunneling splittings have been investigated for a normal-form Hamiltonian whose phase space consists of two copies of the structure, as 
depicted in the left panel of Fig.~\ref{fig:Mouchet}. 
Island-like structures reminiscent of nonlinear resonances appear in the phase space portrait. 
Note, however, that these islands are not generated by a resonance, since the system is one-dimensional. 
Nevertheless, as far as the phase-space topology is concerned, one can say that the setup mimics the situation assumed in the RAT scenario.

Ref.~\cite{le2013semiclassical} revealed that the energy splitting between the two symmetric energy surfaces $\Gamma_{\rm in}$ and $\Gamma'_{\rm in}$ is not governed by the instanton that directly connects $\Gamma_{\rm in}$ and $\Gamma'_{\rm in}$ (the dark blue curve in the figure), but rather by a complex trajectory that first passes through the outer energy surfaces $\Gamma_{\rm out}$ and $\Gamma'_{\rm out}$.
Such a  situation, that is, a tunneling path via the island-like structures dominating over the direct path, 
 is precisely the one assumed in the RAT theory.
It is important to note that the intermediate surfaces $\Gamma_{\rm out}$ and $\Gamma'_{\rm out}$ have the same energy as $\Gamma_{\rm in}$ and $\Gamma'_{\rm in}$, and that there are no complex paths connecting energy surfaces with different energies. 
The same holds for the pendulum Hamiltonian~\eqref{eq:Hres-effective}: instanton trajectories exist only between symmetrically located energy surfaces of equal energy in phase space.

\begin{figure}[htbp]
\centering
\includegraphics[width=0.60\linewidth]{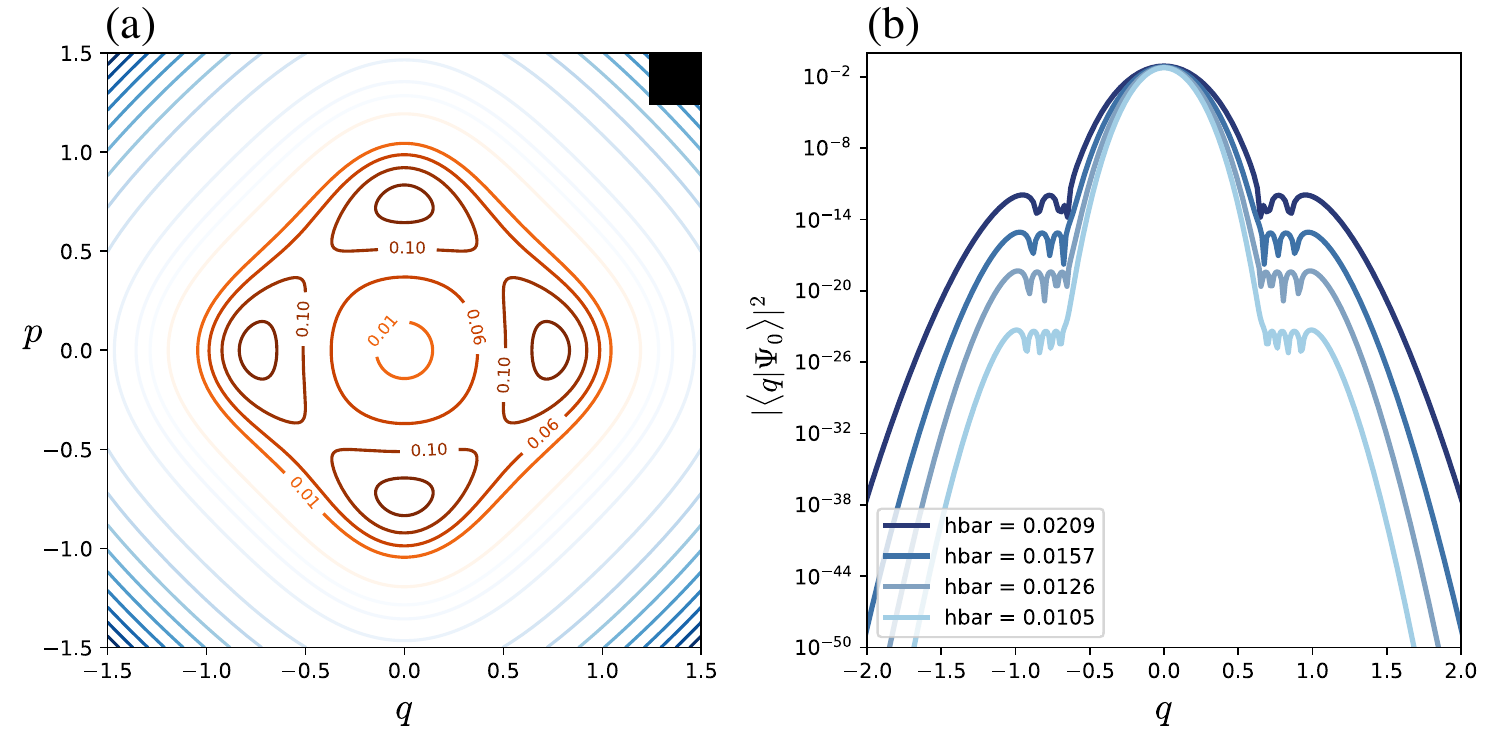}
\caption{
\label{fig:integrable_system}
(a) Classical phase space portrait for the Hamiltonian (\ref{eq:jeremy_normal}). 
(b) The ground state quantum eigenfunction with different values of the Planck constant. 
The black box at the right upper corner in the plot (a) represents the Planck cell for the case of $\hbar=0.0105$. 
Adapted from Ref.~\cite{hanada2015}.
}
\end{figure}

What is even more important is that, in the $1/\hbar$-dependence of the tunneling splitting for the normal-form Hamiltonian, a sequence of spikes due to avoided crossings does appear, but plateaus do not (see the right panel of Fig.~\ref{fig:Mouchet}). 
This is, in a sense, an expected consequence of the fact that the complex-path approximation works. 
For any value of $1/\hbar$, the trajectory passing via the energy surfaces $\Gamma_{\rm out}$ and $\Gamma'_{\rm out}$ always dominates, and the slope of the $1/\hbar$-dependence of the tunneling splitting is determined solely by the imaginary part of the action of the associated complex path.

A one-dimensional system with island-like structures may, at first glance, look like nonlinear resonances in multi-degree-of-freedom systems.  
However, the $\hbar$-dependence of the tunneling tails is fundamentally different.
To see this more clearly, we  consider the following one-degree-of-freedom Hamiltonian~\cite{hanada2015}:
\begin{equation}
  \label{eq:jeremy_normal}
  H(q,p) = H_0(q,p) + \eps H_1(q,p)
\end{equation}
with
\begin{subequations}
  \begin{align}
  \label{eq:jeremy_h0}
    H_0(q,p) &= \tfrac12\,(q^2 + p^2)+ a\,(q^2 + p^2)^2,\\
    H_1(q,p) &= p^4  - 6p^2 q^2 + q^4.
  \end{align}
\end{subequations}
The classical phase space shown in Fig.~\ref{fig:integrable_system}(a) is essentially the same as 
that studied in Ref.~\cite{le2013semiclassical}. 
Figure~\ref{fig:integrable_system}(b) illustrates the ground-state eigenfunction for several values of $\hbar$. 
A plateau appears precisely at the location of the island-like structure in phase space, and one might think that this is precisely the enhancement in tunneling probability envisioned by RAT. 
However, a similar step-like structure of the tunneling tail also appears in one-degree-of-freedom systems with a triple-well potential, and therefore cannot be attributed to nonlinear resonances.
What is important is the $\hbar$-dependence of the tunneling tail.
As for the plateau height, one finds, just as in the above example, only a simple exponential dependence. 
Moreover, the plateau position does not shift even when $\hbar$ is varied.
This stands in sharp contrast to the $\hbar$-dependent plateaus observed in the non-integrable systems, as demonstrated in Subsec.~\ref{sec:ultra_integrable}.

\subsection{Quantum resonance}
\label{sec:QR}

In this Subsection, instead of classical nonlinear resonances discussed in the previous Section, 
we examine the role played by {\it quantum resonances} in understanding the enhancement 
of tunneling probability.
To begin with, let us recall that the system under consideration is a one-dimensional system subject to periodic driving, as in Eq.~(\ref{eq:kicked-ham}). 
As an unperturbed Hamiltonian, we take the truncated BCH Hamiltonian $H_\eff^{(M)}$, 
whose eigenvalue equation is given by (\ref{eq:BCH_eigenstate}). 
Note that $H_\eff^{(M)}$ is a one-dimensional Hamiltonian, and so it is completely integrable. 
Below, we drop the order $M$ of truncation of the BCH series unless otherwise stated. 
Let $\Omega$ be the angular frequency of the external drive,
and define the resonance energies with respect to $E_n$ by 
\begin{equation}\label{eq:qres}
E^{(\mathrm{res},n)}_{k} = E_n + k\hbar \Omega,  ~~(k = 1, 2, \cdots)
\end{equation}
We say that the {\it quantum resonance} occurs between the states $\ket{\,J_n\,}$  and $\ket{\,J_m\,}$ if the condition $E_m = E^{(\mathrm{res}, n)}_{k}$ holds for some $k$. 
Note that the quantum resonance condition does not necessarily have any relation to the classical resonance condition. 
For example, for $s = 1$, the classical resonance condition~(\ref{eq:rs-condition}) gives $ \Omega = r\omega(I) $, where $ r $ runs over positive integers and $ \omega(I) $ varies accordingly. 
However, this does not establish any one-to-one correspondence between $\Omega $ and $ \omega(I)$.

\begin{figure}[htbp]
\centering
\includegraphics[width=0.90\linewidth]{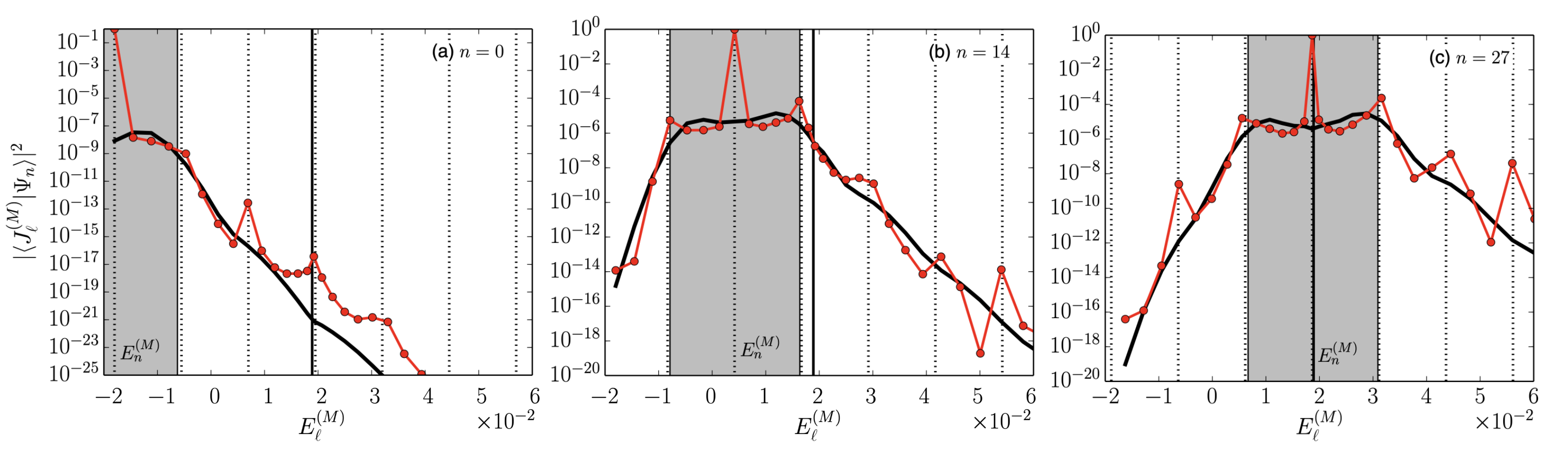}
\caption{
	\label{fig:eigenstate_resonance}
	The red curves represent the eigenstates $\bracket{J_\ell^{(M)}}{\Psi_n}$ in the action representation, 
	plotted as a function of $E^{(M)}_\ell$ 
	for (a) $n=0$, (b) $n=14$ and (c) $n=27$, respectively. 
	The black curves show the matrix elements $\bra{J_\ell^{(M)}}\Delta \hat{U}\ket{J_n^{(M)}}$.
	Here, we used the 7-th order BCH Hamiltonian as the basis $\ket{J_\ell^{(M)}}$.
	The vertical black solid line indicates the separatrix energy, while the dotted lines indicate the energies satisfying $E^{(\mathrm{res},n)}_{k}  = E_{n}^{(M)} + k\hbar \Omega~ (k = 0,1,2,\ldots)$. 
	The Planck constant is set to $h=1/80$. 
	Adapted from Ref.~\cite{hanada2015}. 
}
\end{figure}

Figure~\ref{fig:eigenstate_resonance} presents quasienergy eigenstates expressed in the BCH basis, 
revealing quantum resonances in the tunneling tail of the eigenfunctions.
One notices that peaks emerge at the quantum-resonant positions not only for the ground state ($n=0$) but also for the excited states ($n=14$ and $27$). 
Importantly, as is evident from the definition, the energies at which these peaks occur shift with 
$\hbar$. 
Although we present results for a single fixed $\hbar$ in Fig.~\ref{fig:eigenstate_resonance}, eigenfunctions for different values of $\hbar$ likewise reveal peaks at the quantum-resonant states, with their energies shifting in proportion to $\hbar$. 

In Fig.~\ref{fig:maximal_mode}, we have shown that the staircase structure in the energy splitting is well reproduced by the maximal mode of the contribution spectrum.
The maximal mode introduced there is exactly the state that satisfies the quantum resonance condition.
Hence, one can understand the origin of the staircase as follows:
as $1/\hbar$ increases, a quantum-resonant state located outside the separatrix becomes the maximal mode and approaches the separatrix, resulting in a plateau in the splitting plot.
As $1/\hbar$ increases further, the maximal mode enters the region inside the separatrix; however, it remains dominant for some range of the parameter, producing a decay region after it has entered.
Subsequently, a resonant state of one higher order lying outside the separatrix becomes the maximal mode and forms the next plateau in the tunneling splitting, and this behavior then repeats~\cite{hanada2015}.

\begin{figure}[htbp]
\centering
\includegraphics[width=0.90\linewidth]{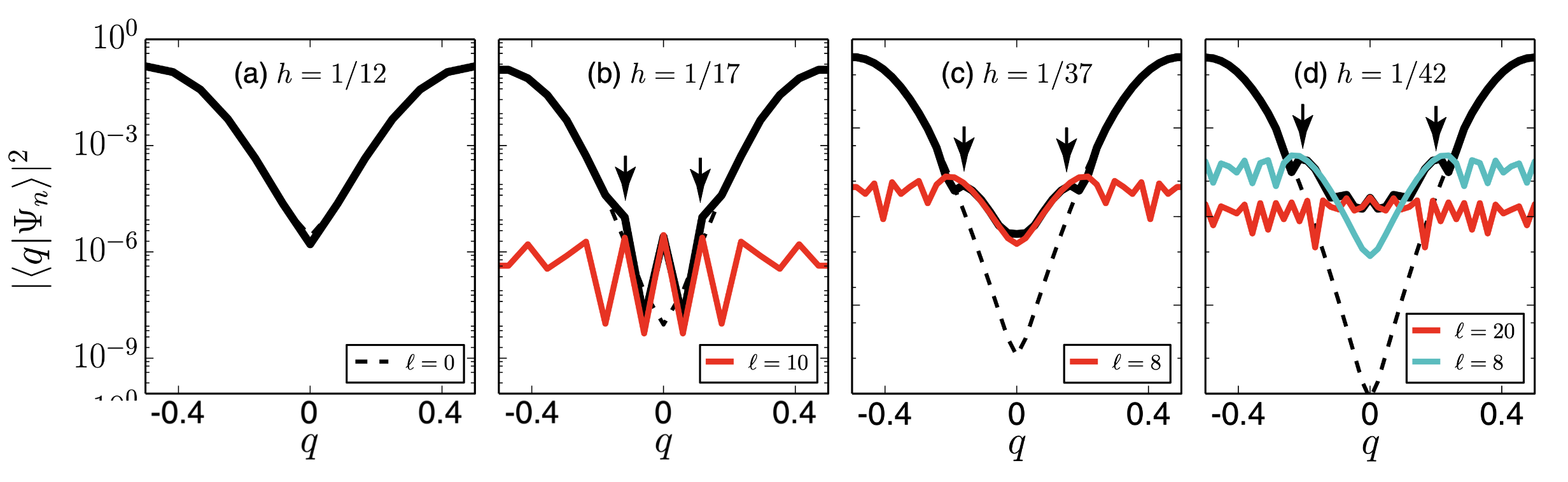}
\caption{
	\label{fig:hump}
	In each panel, the black curve shows the eigenstate $\ket{\Psi_0}$ for $\tau=1$ in (a) the first 
	decay, (b) the first plateau, (c) the second decay, and (d) the second plateau regime, respectively.
	The dashed curve displays the integrable eigenstate $\bracket{\,q\,}{\,J^{(M)}_0}$ at the 
	corresponding $\hbar$ value, and the red and turquoise curves represent the integrable components
	$\bracket{\,q\,}{\,J^{(M)}_\ell\,}\bracket{\,J^{(M)}_\ell}{\,\Psi_0\,}$ at $q=0$, where the value of $
	\ell$ is indicated in each panel. 
	The arrows highlight the positions of the humps.
	Note that the structure around $q=0$ is well reproduced by the maximal mode(s) of the 
	contribution spectrum.	 For further details, see Ref.~\cite{hanada2015}. 
}
\end{figure}

While the maximal mode lies inside the separatrix (the decay region), it shifts as $1/\hbar$ varies. Consequently, as in the energy representation shown in Fig.~\ref{fig:eigenstate_resonance}, 
a hump created by the maximal mode also appears in the tunneling tail of the wavefunction in the coordinate representation, and the position of this hump also shifts with $1/\hbar$ (see Fig.~\ref{fig:hump}). 
Notice, in contrast, that structures supported by invariant structures such as nonlinear resonances in classical phase space do not shift as $\hbar$ is varied.

Quantum resonances arise when an external drive is applied. 
Therefore, the same phenomenon cannot occur in one-dimensional systems. 
This is why the staircase does not appear in the $ 1/\hbar $-dependence of the tunneling splitting (see the right panels of Fig.~\ref{fig:Mouchet} and the result for the normal form Hamiltonian (\ref{eq:jeremy_normal}) in \cite{hanada2015}). 
Moreover, as demonstrated in Fig.~\ref{fig:integrable_system}, the location of the plateau does not exhibit $\hbar$-dependence because it is supported by the island-like structure, which is classically invariant. 
We emphasize that the plateau in the integrable system has a different origin from the hump produced by the maximal mode (see Fig.~\ref{fig:hump}). 

On the other hand, as mentioned in Subsec.~\ref{sec:Avoided_crossings}, although the relationship between avoided crossings and classical resonances is not yet fully understood, some correspondence does exist~\cite{wisniacki2011poincare}. 
Consequently, if the value of $\hbar$ (or some system parameter) takes a particular value such that the energy at which a tunneling doublet undergoes an avoided crossing with a third state also satisfies the quantum resonance condition (\ref{eq:qres}), then at that moment a link is established between classical and quantum resonances. The spike observed at $1/\hbar = 4.5$, discussed above,  is precisely such a moment.
The coupling evaluated by the RAT recipe at this point effectively gives the coupling strength of the quantum-resonant state, i.e., the maximal mode.
Since at that moment $\Omega = 8\omega(I)$ holds, the RAT coupling continues to track the maximal-mode strength as $1/\hbar$ varies. 
This explains why the RAT calculation reproduces the decay of the tunneling splitting with reasonable accuracy (see Fig.~\ref{fig:RAT_test}(a)).

\subsection{Ultra-near-integrable system}
\label{sec:ultra_integrable}

As a model that can more clearly illustrate the role of quantum resonance, we introduce a system extremely close to the integrable limit. 
Not only CAT and RAT but also previous works on dynamical tunneling have studied the situations where classical phase space structures such as chaotic seas, islands of stability, and other kinds of invariant sets in phase space are sufficiently large compared with, or at least comparable to, the size of (effective) Planck cell. 
This is because of our implicit understanding that quantum mechanics cannot resolve classical invariant structures on scales smaller than the (effective) Planck cell. 
In fact, in Subsec.~\ref{sec:Verification_of_RAT}, the $8\!:\!1$ nonlinear resonance employed in the implementation of RAT has a size comparable to that of the (effective) Planck cell. 
In general, however, such phase spaces are highly complex and difficult to analyze even classically, so isolating exponentially small quantum effects under such complicated circumstances is rather challenging.

The so-called \textit{semiclassical eigenfunction hypothesis} concerns eigenstates in the semiclassical limit, conjecturing that, for mixed systems, eigenfunctions become localized exclusively on invariant (regular or chaotic) regions as $\hbar \to 0$~\cite{percival1973}.
As a result, the weight of each state is expected to be proportional to the corresponding phase-space area~\cite{berry1984}.
There is a great deal of evidence supporting this conjecture, provided by directly observing the eigenfunctions for mixed systems~\cite{carlo1998,backer2004,barnett2007}, and the hypothesis is also supported by extensive studies of level statistics~\cite{prosen1993}.
In addition to numerical investigations, rigorous analyses providing further support have also been developed~\cite{marklof2004,galkowski2014,gomes2017,gomes2018}.

It should be recalled, however, that quantum tunneling is an exponentially small effect, whereas the semiclassical expansion in Planck's constant, which underlies the semiclassical eigenfunction hypothesis, is incapable of capturing exponentially small quantities. 
Therefore, it is not surprising to encounter exponentially small effects that are not directly linked to the support of eigenfunctions, even in the semiclassical limit.

The {\it ultra-near-integrable system} introduced in Ref.~\cite{iijima2022quantum} serves to reveal such a subtle nature of the tunneling effect.
It is defined as a class of systems for which the classical invariant structures associated with 
non-integrability are not resolvable in phase space at the scale of the (effective) Planck cell. 
In this sense, the ultra-near integrability is a kind of relative concept and can only be defined through the reference quantum system. 
We now show that ultra-near-integrable systems exhibit behavior that is interesting in its own right, and the quantum resonances discussed above become more pronounced. 

\begin{figure}[htbp]
\centering
\includegraphics[width=0.70\linewidth, trim=0 95mm 0 90mm, clip]{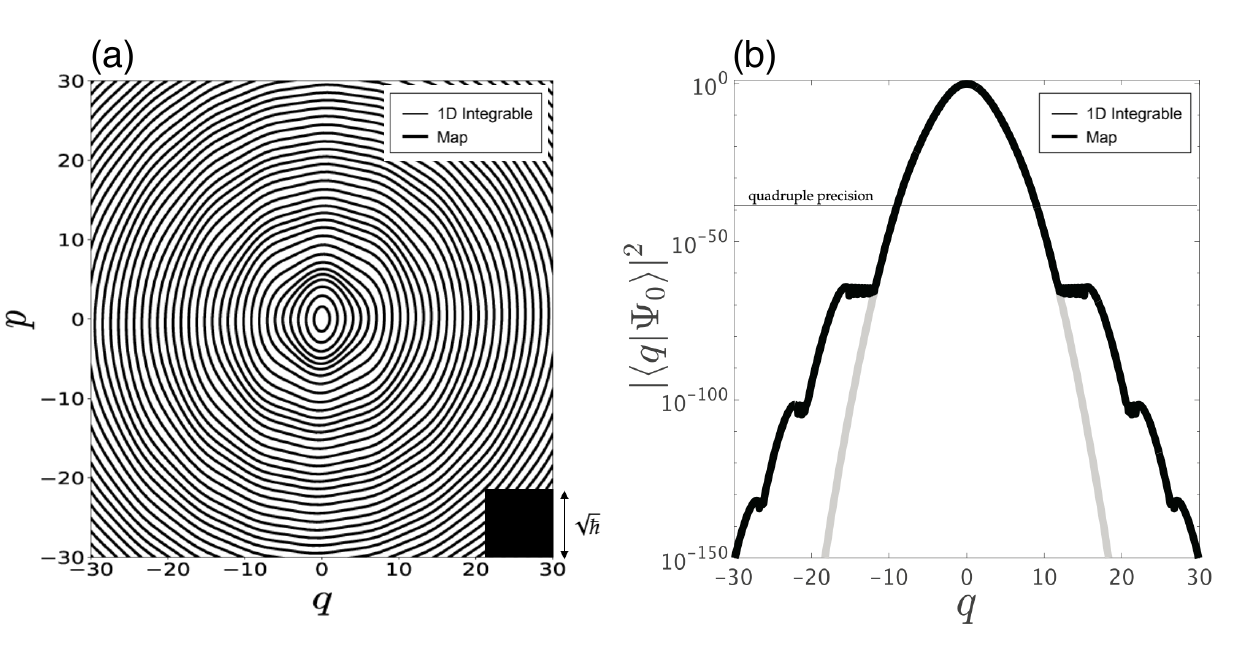}
\caption{\label{fig:ultra-near-integrable}
(a) Phase-space portrait of the classical map (\ref{eq:cmap}) with the potential (\ref{eq:cosine_modulation}).
  The invariant curves of the corresponding one-dimensional system are also shown, but they cannot be distinguished from those of the map.
The parameters of the potential function (\ref{eq:cosine_modulation}) 
are set to $\varepsilon=1.0$, $\tau = 0.05$ and $\lambda=1.2$. 
The black box in the lower-right corner represents the Planck cell with $\hbar =1$. 
(b) The black curve shows the ground state eigenfunction for the quantum map (\ref{eq:qmap_eigen}) with the potential (\ref{eq:cosine_modulation}), and the gray one shows that for the truncated quantum BCH Hamiltonian $\hat{H}_\eff^{(M)}$ with $M=3$. 
The parameters in the potential function (\ref{eq:cosine_modulation}) are set to $\lambda=1.2$ and $\tau = 0.05$, and $\varepsilon=1.0$. The Planck constant is set to $\hbar = 1$. 
}
\end{figure}

For demonstration, we continue to use the kicked-rotor system defined in~\eqref{eq:kicked-ham}, taking the potential in the following form: 
\begin{align}
\label{eq:cosine_modulation}
V(q) =    \frac12 q^2 -2\varepsilon \cos\left(\frac{q}{\lambda}\right) , 
\end{align}
where $ \varepsilon $ and $ \lambda $ denote parameters~\cite{lando2020}. 
In the limit $ \tau \to 0 $, the system tends to the continuous-time Hamiltonian $H(q,p) = T(p) + V(q)$.
Figure~\ref{fig:ultra-near-integrable}(a) shows the classical phase space for the $\tau = 0.05$ case. 
In the lower-right corner of Fig.~\ref{fig:ultra-near-integrable}(a), we indicate the Planck cell used in the quantum calculations below. 
Evidently, at the scale resolvable into Planck cells, one cannot observe any invariant structure that reflects the non-integrability of the system.
Note that Fig.~\ref{fig:ultra-near-integrable}(a) overlays the invariant curves of the one-dimensional system obtained via the BCH approximation, but the two are indistinguishable, at least on this scale.
 
Despite the absence of visible non-integrable structures in the classical phase space, 
the tunneling tail exhibits a staircase structure, as displayed in Fig.~\ref{fig:ultra-near-integrable}(b).
For comparison, we also plot the eigenfunction of the BCH Hamiltonian, 
whose tail decays monotonically and shows no specific features, as expected. 
As we show below, the step structure observed in the ultra-near-integrable system is caused by quantum resonances.

\begin{figure}[htbp]
\centering
\includegraphics[width=0.30\linewidth]{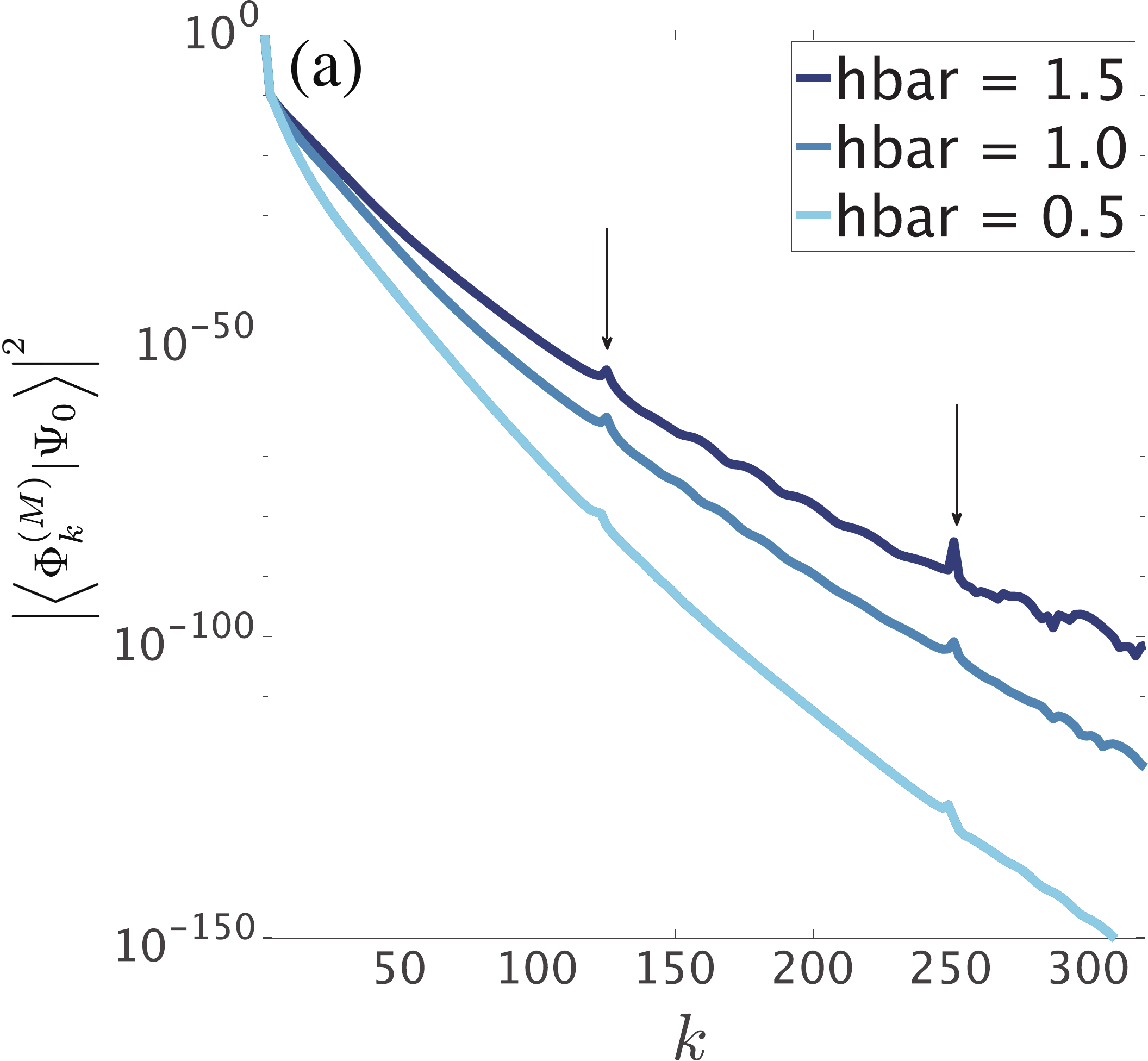}
\raisebox{-1.2mm}{\includegraphics[width=0.323\linewidth]{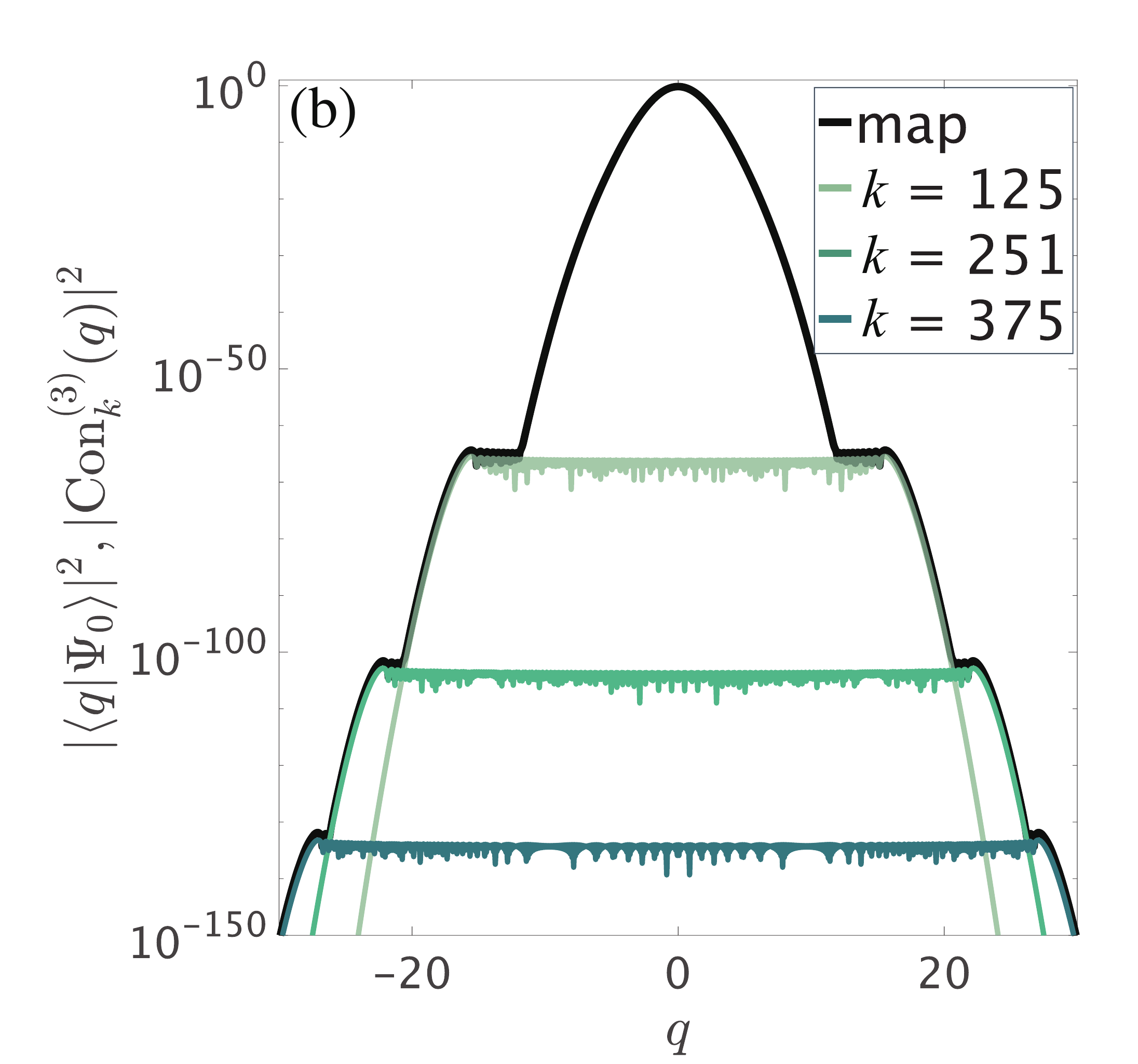}}
\raisebox{-4.1mm}{\includegraphics[width=0.317\linewidth]{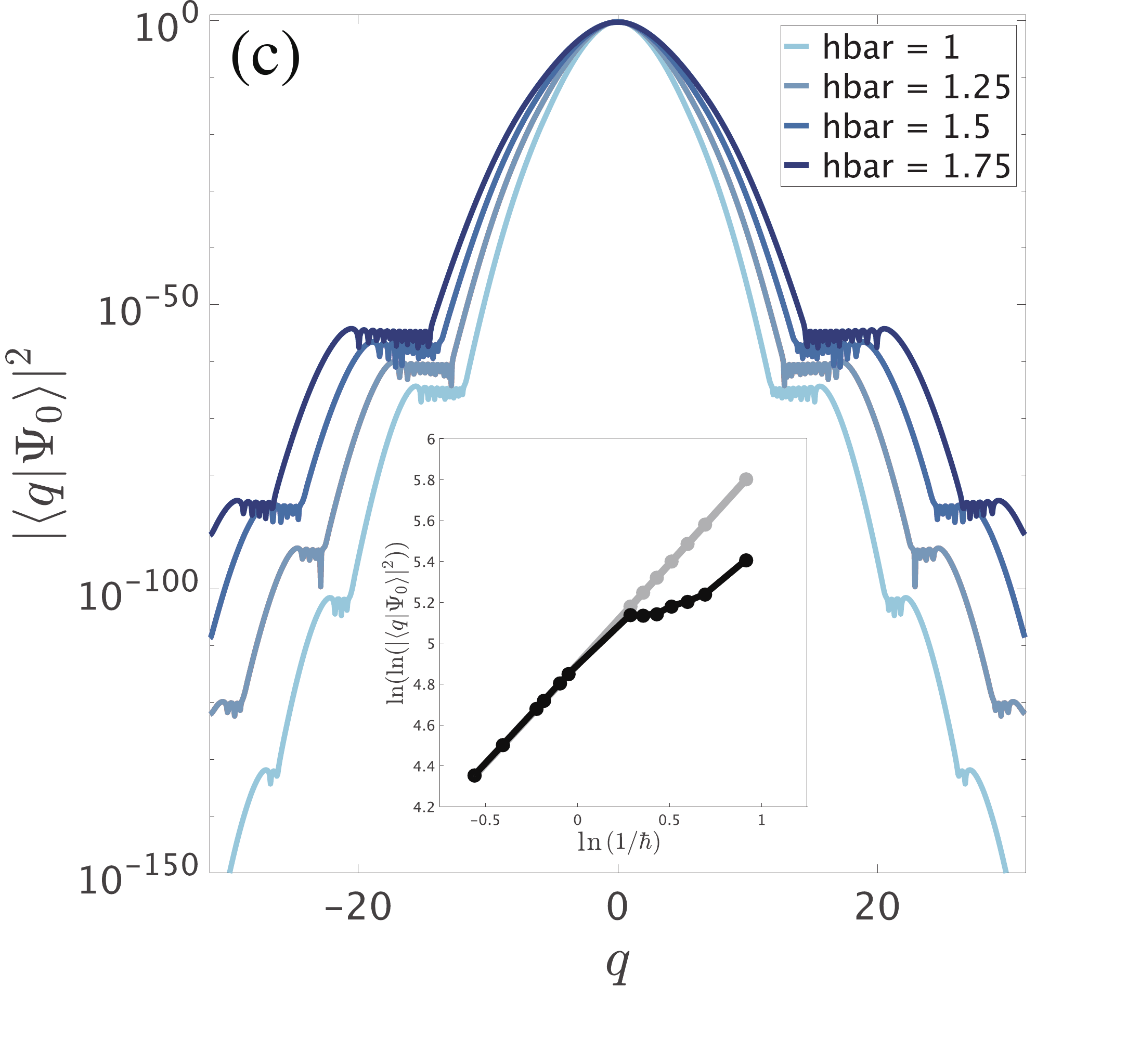}}
\caption{
\label{fig:decomposition}
(a) Ground state eigenfunctions in the quantum BCH representation $|\Phi_k^{(M)} \rangle$ with $M=3$ for different values of $\hbar$. 
(b) Ground state eigenfunction in the $q$-representation in the case $\hbar =1$ (black). 
The states $k=125$, $k=251$, and $k=375$ in the sum (\ref{eq:partical_sum}) are superposed and shown in different green colors.
These states are responsible for the small peaks indicated by the arrows in the plot (a). 
(c) Ground state eigenfunctions for the quantum map (\ref{eq:qmap_eigen}) with 
different values of the Planck constant. 
(Inset) Magnitude of the ground state eigenfunctions at a fixed position ($q=11$) 
plotted as a function of $1/\hbar$.  
The black and gray curves correspond to the ground-state eigenfunction and the truncated quantum BCH Hamiltonian (\ref{eq:bch_hamiltonian}) with $M=3$, respectively.
The parameters are set to $\tau = 0.05$, $\varepsilon=1.0$, and $\lambda = 1.2$. 
Adapted from Ref.~\cite{iijima2022quantum}. 
}
\end{figure}

To this end, we first show in Fig.~\ref{fig:decomposition}(a) the ground state in the quantum BCH basis. 
We can see that the ground state is well approximated by that of the BCH basis, which is manifested by a sharp drop of the curve around $k \sim 0$. 
After the initial drop, the curves decay overall exponentially, except for small peaks indicated by the arrows in the plot. 
The eigenvalues corresponding to the peaks can be read off from Fig.~\ref{fig:decomposition}(a), 
and it exactly satisfies the quantum resonance condition (\ref{eq:qres}). 
Note that Fig.~\ref{fig:decomposition}(a) presents essentially the same plot as Fig.~\ref{fig:eigenstate_resonance}. 
In both figures, quantum resonances appear as spikes.

We can directly confirm that the coupling with the states creating small peaks 
in Fig.~\ref{fig:decomposition}(a) is
responsible for the staircase found in Fig.~\ref{fig:ultra-near-integrable}(b). 
To see this, we expand the 
ground state as \cite{hanada2015}, 
\begin{align}
\label{eq:partical_sum}
\braket{q}{\Psi_0} = \sum_n \mathrm{Con}_n^{(M)}(q), 
\end{align}
where 
\begin{align}
\mathrm{Con}_n^{(M)}(q) := \braket{q}{J_n^{(M)}}\braket{J_n^{(M)}}{\Psi_0}. 
\end{align}
Instead of summing over all states $k$, we retain only those that generate the peaks observed in Fig.~\ref{fig:decomposition}(a). 
As shown in Fig.~\ref{fig:decomposition}(b), the resulting state reproduces the observed staircase.

More importantly, as shown in Fig.~\ref{fig:decomposition}(c), the staircase shifts with $\hbar$. 
This contrasts with the one-dimensional case, shown in Fig.~\ref{fig:integrable_system}(b), where the staircase position remains unchanged as $\hbar$ is varied. 
The shift of the staircase is the same phenomenon as the shift of the hump in the tunneling tail of the standard map found in Fig.~\ref{fig:hump}. 
These effects occur because the energies at which quantum resonances take place vary with $\hbar$. 
This $\hbar$-dependence shown in Fig.~\ref{fig:decomposition}(c) is another piece of evidence that these structures do not originate from specific classical nonlinear resonances.

Observing the wavefunction amplitude at a fixed position reveals a nontrivial $\hbar$ dependence.
As illustrated in the inset of Fig.~\ref{fig:decomposition}(c), the wavefunction amplitude for the BCH Hamiltonian exhibits a simple exponential decay, as expected.
On the other hand, the amplitude for the quantum map shows an exponential decay in a large $\hbar$-regime, but it switches to a stretched exponential-type decay. 
The slope again returns to 1 for a smaller $\hbar$ regime. 
Notice that stretched exponential regions appear when the observation point hits a plateau region of the wavefunction. 
Due to computational limitations, we cannot access smaller values of $\hbar$, but it is reasonable to expect that this staircase-like structure will persist as $\hbar$ is further reduced, since the regular staircase structure induced by quantum resonances extends deep into the tunneling tail.

The stretched exponential decay region has the same origin as the plateau region observed in the $\Delta E$ vs. $1/\hbar$ plot for the standard map (see Fig.~\ref{fig:maximal_mode}). 
The reason is that, in the latter case, the plateau reflects the contribution at $q=0$ from the real invariant manifold associated with rotational motion, whereas in the present case, the stretched-exponential decay similarly reflects the real invariant manifold of the quantum-resonant excited state.
We also note that even for the standard map with sufficiently small $\tau$, in which no visible nonlinear resonances resolvable at the scale of a Planck cell in phase space appear, the $\Delta E$ vs. $1/\hbar$ plot exhibits a staircase structure~\cite{hanada2015}.

Finally, we revisit the role of classical nonlinear resonances. As mentioned in Subsec.~\ref{sec:RAT}, a spike appears precisely at the point where the plateau region switches to the decay region; at that moment, the ground-state doublet undergoes an avoided crossing with a third state. 
The tunneling process there can be modeled by the local integrable Hamiltonian, 
but this does not by itself justify the claim that, for other values of $1/\hbar$, in particular, across the entire second decay region, the classical resonance responsible for the spike 
continues to mediate tunneling.
Establishing this point would need a direct investigation based on a fully semiclassical analysis, 
instead of a hybrid computation employing quantum perturbation theory. 
Any such analysis must be based on the non-integrability of the system because, as we have seen earlier, the $\Delta E$ vs.$1/\hbar$ plot for the one-dimensional normal-form Hamiltonian does not exhibit a staircase structure.
In any case, it would be important to recognize that the appearance of spikes and persistent enhancement are distinct phenomena.

\section{Complex paths and quantum tunneling}
\label{sec:Complex_path}


\subsection{Complex semiclassical approach to continuous-time systems}
\label{sec:Instanton_path}

As quantum tunneling lacks a counterpart in classical mechanics, it is necessary to extend classical dynamics into the complex plane to explore a classical interpretation of tunneling.
A semiclassical theory based solely on real classical trajectories cannot approximate quantum tunneling, no matter how far one pushes the expansion in $\hbar$, since, as we mentioned above, tunneling is an exponentially small effect and hence intrinsically non-perturbative.

The importance of complex classical dynamics in tunneling problems has long been recognized, and descriptions of tunneling in terms of complex classical trajectories were, as briefly mentioned in Introduction,  already widespread well before studies of tunneling in chaotic systems began. 
The most familiar complex path describing quantum tunneling is the so-called \textit{instanton}---a classical trajectory that crosses an energy barrier along a contour in complex time---discovered independently in quantum field theory and chemical reaction theory~\cite{george1972complex,miller1974quantum,callan1977fate,coleman1977fate}.
In this connection, in the perturbative analysis of coupled anharmonic oscillators, it was shown that the large-order behavior of the perturbation coefficients is governed by the instanton action~\cite{banks1973coupled1,banks1973coupled2}. 
This can be regarded as  a precursor to what is nowadays known as resurgence theory, a mathematical framework for treating exponentially small effects or equivalently non-perturbative effects~\cite{dingle1973,ecalle1981,voros1983,olver1997asymptotics,delabaere1997,kawai2005,balser2006divergent,mitschi2016divergent,aniceto2019primer}. 

In one-dimensional systems, it is well established that an instanton-based description of tunneling applies not only to systems with double-well or metastable potential but much more broadly~\cite{le2010instantons,le2013,harada2017riemann}. 
The complexified constant-energy curve is realized as a higher-genus Riemann surface, and action integrals are obtained by integrating along a basis of independent closed (homology) cycles on that surface~\cite{harada2017riemann}. 
Motivated by recent developments in resurgent theory, 
Ref.~\cite{tanizaki2014real} has reformulated the real-time Feynman path integral using complex analysis and Picard--Lefschetz theory, a complexified version of Morse theory.
The ordinary real-time path integral suffers from poor convergence because of the rapid oscillations of the phase factor $iS/\hbar$, making it difficult to handle both mathematically and numerically.
The authors in Ref.~\cite{tanizaki2014real} address this problem by complexifying the integration contour and deforming it into a sum over Lefschetz thimbles.

In multi-dimensional systems, as long as the dynamics is integrable, the complexified energy surface remains analytic; accordingly, action integrals taken over the complexified energy manifold determine the tunneling splittings~\cite{creagh1994tunnelling}. 
We note rigorous mathematical results claiming that the tunneling splitting $\lvert E_1 - E_0 \rvert$ satisfies
\begin{equation}
\lim_{\hbar \to 0} \Bigl( -\hbar \ln \lvert E_1 - E_0 \rvert \Bigr) = \rho(a,b),
\label{eq:Simon}
\end{equation}
where $E_0$ and $E_1$ are the ground- and first-excited-state energy eigenvalues of a system with a double-well potential, and $\rho(a,b)$ is the so-called Agmon distance between the well minima $a$ and $b$~\cite{simon1983,simon1984}.
The Agmon distance corresponds to the (Euclidean) action associated with the instanton trajectory.
Importantly, this estimate holds not only in one-dimensional but also in multi-dimensional systems.

In non-integrable systems, the complexified energy surface on which the instanton propagates is generally destroyed, and one accordingly expects a new type of tunneling transport to emerge~\cite{Creagh98}.
Wilkinson's pioneering work analyzed tunneling from a WKB perspective in situations where the analyticity of the energy surface is lost, i.e., when the system becomes non-integrable~\cite{wilkinson1986tunnelling,wilkinson1987multidimensional}.
The setting considered there is a two-degree-of-freedom system with double-well potential, where the energy surface is analytically continued from each well into the classically forbidden region. 
In the absence of coupling between the two degrees of freedom, the system is completely integrable, consisting of two independent one-dimensional subsystems, the complex action along the instanton path then yields the tunneling splitting. 
In the presence of coupling, the paper assumes that the left and right complex tori do not merge smoothly into a single torus, but instead intersect transversely at a point with a finite angle~\cite{wilkinson1987multidimensional}. 
Under this hypothesis, Wilkinson derived a formula for the tunneling splitting~\cite{wilkinson1987multidimensional}, and this idea has subsequently been applied to a model describing chemical reactions~\cite{takada1995effects,takada1994wentzel}.

It is far from obvious whether the loss of analyticity of the underlying energy surface immediately leads to the disappearance of instanton trajectories. 
To the best of the author's knowledge, there is no general theory that answers this question, and so 
we have no choice but to deal with each case individually. 
To address this issue, it may be easier to analyze scattering systems rather than bound systems such as the double well. 
Creagh et al. formulated an instanton description of barrier crossing for two-dimensional scattering systems in terms of the scattering matrix and carried out a detailed analysis~\cite{creagh2004classical,creagh2005semiclassical,drew2005uniform}. 
In general, when one considers scattering in two or more dimensions, unlike the one-dimensional case, the incident wave possesses transverse degrees of freedom in addition to its propagation direction. 
By employing methods based on the scattering matrix or the Green's function, one can separate the tunneling process from the preparation of the incident wave, thereby enabling the treatment of scattering even when the incident wave resides in a chaotic potential well and forms quasi-bound states. 
Even in such situations, it was shown that the tunneling probability is maximized when the semiclassically prepared incident wave has significant overlap with a neighborhood of the particular trajectory that crosses the barrier with the minimal imaginary action. 
In particular, it is important to note that this optimal trajectory can be regarded as a generalization of the instanton that describes tunneling in integrable systems.
In other words, in such a scattering setting the instanton survives; however, its role depends on the energy of the incident wave and on the strength of the perturbation added to an otherwise integrable scattering potential. 

The role of the instanton in a simple scattering system under periodic driving, as well as a new tunneling mechanism, has been explored in detail in a series of works by Takahashi and Ikeda~\cite{takahashi2000complex,takahashi2003complex,takahashi2005intrinsic,takahashi2006anomalously}.
In their setting, the stable and unstable manifolds emanating from a fixed point at the center of the potential are distorted through stretching and folding. However, because the stable and unstable manifolds do not intersect in real phase space, chaos does not  appear in the real plane. 
Even so, because the singularities in the complex plane are of a different nature from those in completely integrable systems, the resulting semiclassical wave exhibits behavior distinct from that in integrable systems.
They found that when the amplitude of the periodic driving is sufficiently weak, the instanton trajectory governs the tunneling process, whereas once the amplitude exceeds a certain threshold, which is determined by the imaginary part of a singularity in complex space, referred to as the {\it critical point}, orbits guided by the stable and unstable manifolds in the complex domain become the primary contributors to the tunneling probability. 
Incidentally, a similar mechanism was identified by Levkov et al., who refer to the complex trajectory that replaces the instanton as a {\it sphaleron}~\cite{levkov2007complex,levkov2007unstable,levkov2009signatures}.

\begin{figure}[htbp]
\centering
\includegraphics[width=0.90\linewidth, trim=0 110mm 0 110mm, clip]{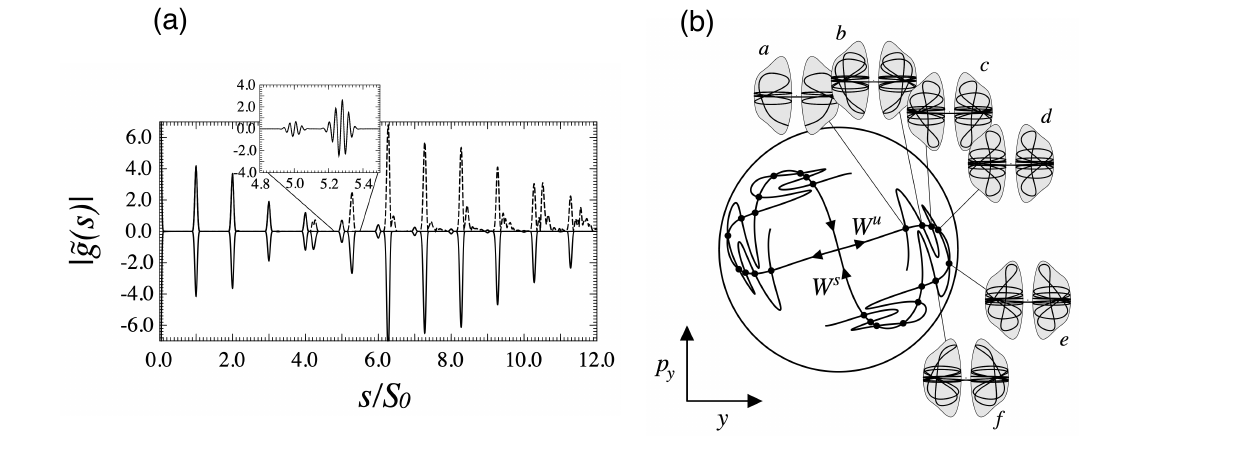}
\caption{
\label{fig:Creagh_Whelan}
(a) Dashed curve: Fourier transform of $g(q)$ obtained quantum mechanically.
Upper solid curve: semiclassical prediction using just the axial orbits.
Lower solid curve: theory, using the six homoclinic families shown in (b).
Inset: the quantum-mechanical Re$\,\tilde{g}(s)$ and the semiclassical prediction in
a limited range (note that there are two superimposed curves).
Peaks not accounted for by the lower curve correspond to non-computed secondary intersections.
(b) 
The stable and unstable manifolds, $W^u$ and $W^s$, intersect to form homoclinic trajectories; six distinct trajectories are labeled (a)-(f). 
For each case, we plot, in configuration space, the trajectory of the corresponding tunneling orbit. The orbit shown is the full periodic orbit, obtained as the double iterate of the pseudo-periodic orbit used in the calculation. Only the real parts are displayed; the imaginary components are too small to be visible.
Reproduced from Ref.~\cite{creagh1999homoclinic}.
}
\end{figure}

Regarding the influence of chaos in real phase space on quantum tunneling, 
Creagh and Whelan developed a semiclassical complex periodic-orbit theory that explains the fluctuations of the tunneling splitting in a double-well system, where the dynamics inside each well becomes fully chaotic~\cite{creagh1996complex,creagh1999homoclinic}.
Rather than treating each splitting $\Delta E_n = E_n^+ - E_n^-$ individually, they introduced the density of states weighted by the splitting, 
\begin{eqnarray}
f(q): =\sum_n \Delta q_n\,\delta\!\left(q-q_n\right). 
\label{eq:tunneling_function}
\end{eqnarray}

Furthermore, instead of dealing with each energy level $E_n$, they ask for the value $q_n$ for which a given energy $E$ is an eigenvalue by regarding $q=1/\hbar$ as a parameter, allowing us to focus on a fixed classical dynamics.  
For this quantity, they derived a semiclassical formula in a form analogous to Gutzwiller's trace formula~\cite{gutzwiller1971periodic},
\begin{eqnarray}
f(q)\approx f_{0}(q)+\frac{1}{2\pi}\,\mathrm{Re}\sum_{r=1}^{\infty}\sum_{\gamma \,\in \,P_{r}} A_{\gamma}\,e^{i q S_{\gamma}}. 
\label{eq:Creagh_formula}
\end{eqnarray}
The sum is taken over complex periodic orbits, and a trick to make the search of periodic orbits possible is to observe the dynamics on an appropriately chosen surface-of-section. 
On such a  section, the periodic orbits are expressed in the form $F^{r}\mathcal{F}$, where $F$ denotes the first-return map on the real plane and $\mathcal{F}$ is the complex map describing the tunneling transition over the energy barrier.
The search for periodic orbits is carried out in the vicinity of a complex orbit with the minimal imaginary action and it is taken as the reference orbit.
The function $f_0(q)$ in the sum (\ref{eq:Creagh_formula}) is determined by this reference orbit and sets the average behavior of the tunneling splitting, appearing as the periodic spikes in Fig.~\ref{fig:Creagh_Whelan}.
On the other hand, to account for deviations from harmonic peaks, it is necessary to incorporate 
other complex orbits, which are supposed to satisfy two requirements: their imaginary action should be close to that of the reference orbit, and they should ``explore'' the entire phase space so as to encode information about the spectrum of tunneling splittings as a whole.
They found that homoclinic orbits are candidates that satisfy both requirements and are therefore included in the sum~(\ref{eq:Creagh_formula}) (see also Fig.~\ref{fig:Creagh_Whelan}).

It is interesting to see that the behavior of the complex orbits incorporating homoclinic orbits introduced there are similar to that predicted by the theory of complex dynamics, which will be explained in the following Subsections. 
As emphasized below, in a mixed phase space, the complex orbits that dominate the semiclassical contributions follow regular motion along complexified KAM tori as long as they stay within the regular region.
Once they enter the chaotic region, however, they approach the real plane and behave almost like real chaotic orbits. Similarly in the case illustrated in Fig.~\ref{fig:Creagh_Whelan}, the homoclinic orbits exploring the chaotic region are not strictly real, but a quasi-real trajectory with only a very small imaginary component, as pointed out in the caption of Fig.~\ref{fig:Creagh_Whelan}.

\subsection{Time-domain semiclassical approach to quantum maps}
\label{sec:Time-domain}

In what follows, we introduce a complex semiclassical approach to dynamical tunneling using 
the quantum map (\ref{eq:qmap}).
In the real dynamics, the associated classical map (\ref{eq:cmap}) can be regarded as a model for the Poincar\'e map of the continuous Hamiltonian system. 
Hence, features of the continuous-time dynamics are expected to carry over to the discrete map.
On the other hand, once one complexifies a continuous-time system, this correspondence is no longer obvious.
This is because, in the continuous-time case, one must complexify not only the dynamical variables $(q,p$) but also time $t$, making the situation more complicated. 
However, aside from studies that, for example, discuss integrability in terms of the nature of singularities in the complex-time plane~\cite{bountis1982integrable,ruiz1999differential,morales2001galoisian}, studies 
on complex classical mechanics with complex time remain limited.
The analysis can become involved, as found in  Refs.~\cite{takahashi2000complex,takahashi2003complex,takahashi2005intrinsic,takahashi2006anomalously}, and the correspondence with discrete maps may fail in some cases. 

For the semiclassical analysis of the symplectic map,
we first present a time-domain semiclassical framework for describing dynamical tunneling.
The quantum time evolution of maps is governed by a discrete analogue of the Feynman-type propagator,
\begin{eqnarray}
\label{propapator_general}
\langle {\,\cal B}\, |\, U^n \,|\, {\cal A} \,\rangle
= \int \cdots \int \prod_j dq_j \prod
_jdp_j
\exp \left( \frac{i}{\hbar}S_n({\cal A}, {\cal B})\right). 
\end{eqnarray}
Here, we denote the initial and final states symbolically by ${\cal A}$ and ${\cal B}$.
These may represent a coordinate $q$, a momentum $p$, or an action $I$, but this choice does not affect the following argument. 
The function $S_n({\cal A}, {\cal B})$ is the classical action, defined so that the original mapping relation is recovered by imposing the variational condition. 
An advantage of the time-domain semiclassical approach is that it yields detailed insight into 
the time evolution of wavefunctions through complex classical trajectories.
On the other hand, because the argument inevitably depends on the representation one chooses, it cannot lead to a closed semiclassical expression of the kind proposed in Refs.~\cite{creagh1996complex,creagh1999homoclinic}.

The semiclassical approximation is obtained by evaluating the multiple integral
$\langle {\cal B}\,|\,U^n\,|\,{\cal A}\rangle$
using the saddle-point method. 
The resulting expression is known as the Van Vleck-Gutzwiller propagator, which takes the form, 
\begin{eqnarray}\label{eq:smprop}
\langle {\cal B}\,|\,U^n\,|\,{\cal A}\rangle
\simeq \sum_{\gamma} A_n^{(\gamma)}({\cal A},{\cal B})\,
\exp\left(
\frac{i}{\hbar}S_n^{(\gamma)}({\cal A},{\cal B})
+ {\rm i}\frac{\pi}{2}\mu^{(\gamma)}
\right),
\end{eqnarray}
where $A_n^{(\gamma)}({\cal A},{\cal B})$, $S_n^{(\gamma)}({\cal A},{\cal B})$, and $\mu^{(\gamma)}$
denote, respectively, the amplitude factor determined by the stability of each classical orbit $\gamma$,
the corresponding classical action, and the Maslov index.
The semiclassical propagator is thus obtained by summing over all classical trajectories $\gamma$ that begin at ${\cal A} = \alpha$ and end at ${\cal B} = \beta$.
Note that initial and final values $\alpha$ and $\beta$ should be 
real-valued since they are observables. 
The set of points contributing to the semiclassical sum 
is thus expressed in general as
\begin{eqnarray}\label{M-set}
\M_{\,n}^{\,\alpha,\,\beta} = \I \cap F^{-n}(\F), 
\end{eqnarray}
where 
\begin{eqnarray}\label{cl-manifold1}
\I = \{ (p,q) \in {\Bbb C}^2 \,| \, {\cal A} = \alpha  \in \bR  \},  
\h{5mm}
\F = \{ (p,q) \in {\Bbb C}^2 \, | \, {\cal B} = \beta \in \bR  \}. 
\end{eqnarray}

\subsection{Complex path decomposition of wavefunction}
\label{sec:Complex_path_decomposition}

\begin{figure}[htbp]
\centering
\includegraphics[width=0.90\linewidth, trim=0 05mm 0 05mm,clip]{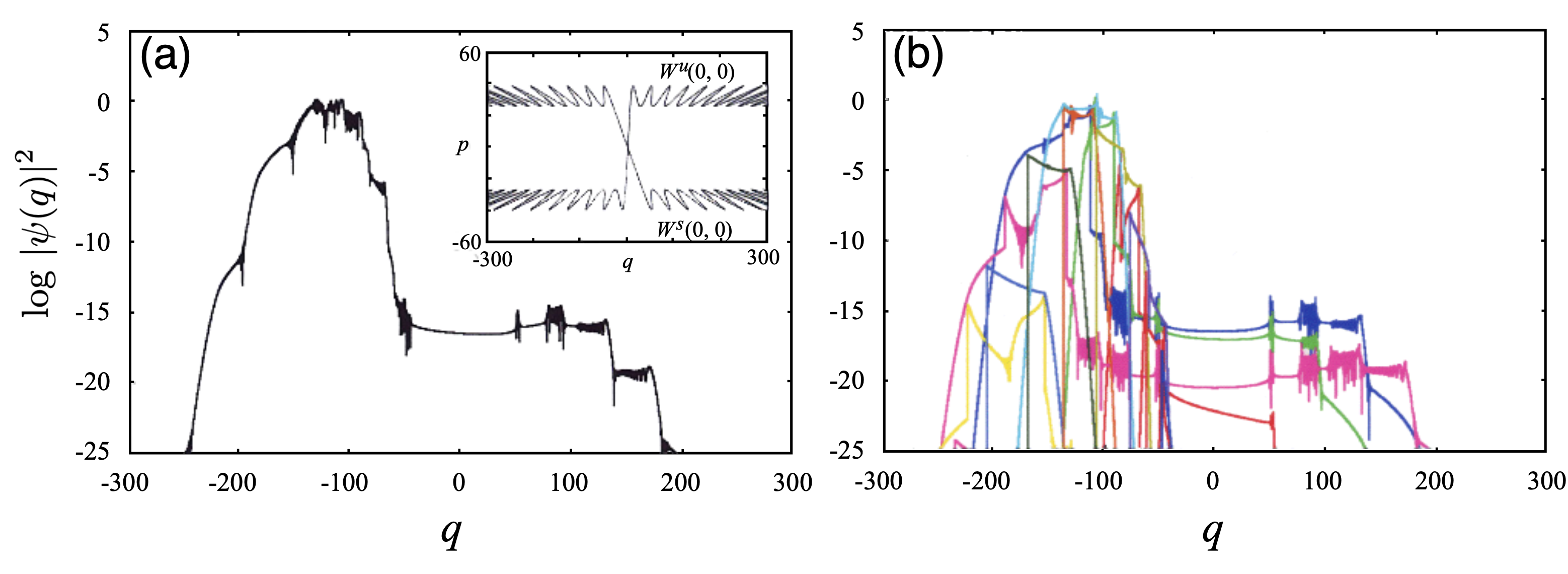}
\caption{
\label{fig:Onishi}
(a) Quantum and (b) semiclassical wavefunctions $\bigl| \langle q | U ^n | \Psi \rangle \bigr| ^2$ 
for the quantum map~(\ref{eq:qmap}) with potential $V(q) = k \exp \left( \gamma q^2 \right)$.
The incident wave packet is a minimum wave packet placed sufficiently far from the center of the potential at $q=0$, on the side $q<0$. 
These figures illustrate the moment shortly after the center of the wave packet has reached $q=0$. 
The larger-amplitude part on the left represents the reflected wave, while the part extending to the right shows the transmitted wave. 
Since the initial energy is far below the barrier height, the transmitted wave arises from quantum tunneling.
(Inset) $ {\cal W}^s(0,0)$ and
$ {\cal W}^u(0,0)$ represent stable and unstable manifolds for the fixed point at $(q,p)=(0,0)$ in the real plane. 
The semiclassical wavefunction is obtained by the superposition of different colored components, 
each of which is associated with a chain-shaped structure, as shown in Fig.~\ref{fig:M-set}. 
The chain-shaped structure itself is composed of many complex trajectories (see the text). 
For further details, see Ref.~\cite{onishi2001tunneling}. 
}
\end{figure}

Using a scattering map as an example, we demonstrate that calculations based on a semiclassical propagator incorporating complex trajectories reproduce the time evolution of the quantum system, including the tunneling tail, accurately down to fine details~\cite{onishi2001tunneling,onishi2003semiclassical}.
The scattering map considered here is the discrete counterpart of the continuous-time model employed in \cite{takahashi2000complex,takahashi2003complex,takahashi2005intrinsic,takahashi2006anomalously}.
The system under consideration is, as before, the discrete map (\ref{eq:cmap}), for which we take the potential
\begin{eqnarray}
V(q) = k \exp (-\gamma q^2 ), 
\end{eqnarray}
where $k, \gamma >0$ are suitable parameters.
As an initial condition, we take a minimal wave packet of energy far below the potential height, located at $q \ll -1$ as the incident wave. It collides with the potential at the origin, and part is reflected to $q < 0$, while the rest is transmitted to $q > 0$. Figure~\ref{fig:Onishi}(a) presents a snapshot of the wavefunction after this splitting into reflected and transmitted parts.
In the inset of Fig.~\ref{fig:Onishi}(a) we show the stable and unstable manifolds emanating from the fixed point at the potential top. 
As in the continuous-time case~\cite{takahashi2000complex,takahashi2003complex,takahashi2005intrinsic,takahashi2006anomalously}, the stable and unstable manifolds do not intersect in the real plane, so chaos in the strict sense does not appear there; however, in the complex plane, intersections of stable and unstable manifolds take place, giving rise to genuine chaos.
Moreover, reflecting the fact that the potential is a transcendental function, infinitely many complex trajectories arise in the semiclassical sum (\ref{eq:smprop}) even after a finite number of steps.
For this system, one can systematically identify the dominant complex trajectories among infinitely many candidates by exploiting the hierarchical structure of the contributing complex trajectories together with the technique of symbolic dynamics~\cite{onishi2003semiclassical}.
As seen in Fig.~\ref{fig:Onishi}(b), the resulting semiclassical sum (\ref{eq:smprop}) reproduces the exact quantum result remarkably well.
Note that each color-coded semiclassical wave is itself a superposition of many contributing complex trajectories~\cite{onishi2001tunneling,onishi2003semiclassical}. 
The tunneling contribution to the wavefunction thus arises as a superposition of semiclassical waves, each of which is a lumped superposition of numerous complex trajectories, with other groups (indicated by different colors). 
This is in sharp contrast to the case in which the tunneling contribution is fully accounted for by the instanton trajectory alone.

\begin{figure}[htbp]
\centering
\includegraphics[trim = 0 10cm 0 11cm, clip, width=1.00\linewidth]{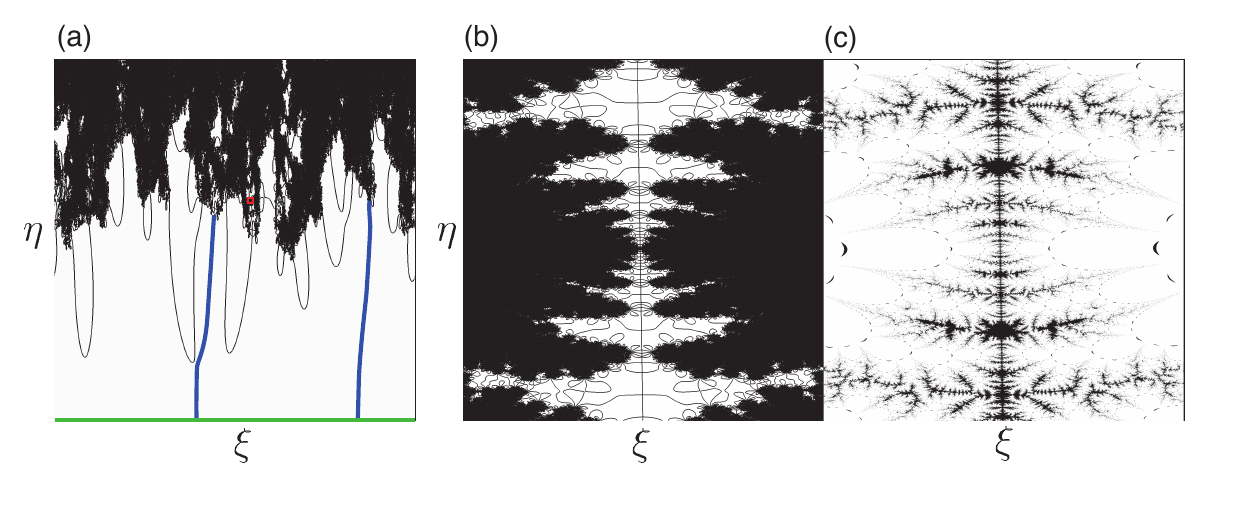}
\caption{
\label{fig:M-set}
A set of initial points that contribute to the semiclassical propagator (\ref{eq:smprop}) 
in the $p$-representation, i.e., ${\cal A} = p$ and ${\cal B} = p'$. 
(a) The case with $T(p) = 
p^2(p/p_d)^6/2\left( (p/p_d)^6+1\right)
+\omega p, ~V(q)=K\sin q$, and 
(b) the case with $T(p) = p^2, ~V(q)=K\sin q$ 
are shown, respectively. 
Panel (b) magnifies a tiny region of the full initial set.
(c) The slice of $K^+$ by $\{p = \alpha \}$, which is obtained by plotting 
the initial points whose trajectories remain within a ball in ${\Bbb C}^2$ 
with a certain sufficiently large radius. 
For further details, see \cite{shudo1998,shudo2002}. 
}
\end{figure}

To identify which complex trajectories among the many candidates actually contribute, we present below another example examined in the $p$-representation, i.e., $\langle p' | U^n | p \rangle$. 
We here visualize the contributing complex paths by displaying the set
\begin{eqnarray}\label{mset}
\M_{\,n}^{\,\ast,\,\beta}  := \bigcup_{\beta \in {\Bbb R}} \M_{\,n}^{\,\alpha,\,\beta} 
= \bigcup_{\beta \in {\Bbb R}} \, \{\,(\xi,\eta)\in\mathbb{R}^2 \,|\, p'(q=\xi+i\eta,\; p=\alpha)=\beta\,\}
\end{eqnarray}
\noindent
on the $q_0$-plane of the slice $\{p_0=\alpha \}$ for some initial condition $\alpha \in {\Bbb R}$.
The set $\M_{\,n}$ on the $q_0 = \xi + {\rm i} \eta$ plane, 
which usually looks like clouds or wisteria trellis on
a macroscopic scale (see Fig.~\ref{fig:M-set}(a)), is decomposed into finer
structures as it is magnified (see Fig.~\ref{fig:M-set}(b)). 
One can see that its basic element is a string with various scales.
Each string represents a trajectory $\gamma$ appearing 
in the semiclassical sum (\ref{eq:smprop}). 
We note that for the one-step semiclassical propagator, i.e., $n=1$ in (\ref{eq:smprop}), 
only the branches connecting with the real axis (shown in green in Fig.~\ref{fig:M-set}(a)) appear, whereas all other complicated structures do not.
Such branches yield a simple, monotonic decay in the tunneling tail and can be regarded as analogs of instanton paths in continuous-time systems.
The saddle point solution obtained in the scattering matrix approach introduced in Subsec.~\ref{sec:Scattering} corresponds exactly to these branches (see the inset of Fig.~\ref{fig:Frischat_Doron}). 
On the other hand, it is not the complex trajectories attached to the real axis but rather those that appear to ``float'' in the complex domain that are responsible for CAT. 

As our starting point, therefore, we have to keep in mind that an enormous number of 
complex paths potentially exist in the semiclassical sum, yet not all of them contribute equally. 
Among all the possible candidates, it was found in~\cite{shudo1995,shudo1998} that the complex orbits hidden in a well-recognizable fractal structure, as presented in Fig.~\ref{fig:M-set}(b), dominate the tunneling contribution. 
Such a structure runs vertically in the initial value plane $(\xi,\eta)$ and is clearly discernible from the other aggregated strings.
As we shall see in the next Subsection, this fractal structure is in fact nothing other than the Julia set of the complex dynamics. 
The Julia set is generally an object in $\mathbb{C}^2$, but here we restrict ourselves to the slice shown in Fig.~\ref{fig:M-set}, obtained by intersecting the Julia set with the manifold specified by the initial condition.

\subsection{Julia sets in the complex dynamics}
\label{sec:Julia_set}

As shown in the previous Section, non-integrability implies that semiclassical theory may involve contributions from a potentially vast number of complex trajectories. 
The question we address here is whether the dominant contribution comes from a single trajectory, from multiple trajectories, or instead from a very large set of complex trajectories whose contributions are comparable in magnitude.
Below, we take two steps toward narrowing down these possibilities. 

Before proceeding, we provide a general classification of trajectories in complex dynamical systems, together with a summary of fundamental results, particularly in the multi-dimensional setting.
Here, we present rigorous results for the H\'enon map~\cite{henon2004two}, which in its standard form is given by 
\begin{eqnarray}
\label{henon}
\displaystyle
\H:
\left(
\begin{array}{c}
x\\
y
\end{array}
\right)
\mapsto
\left(
\begin{array}{c}
y\\
y^2 - x + a
\end{array}
\right).
\end{eqnarray}
The H\'enon map is known to be the simplest two-dimensional polynomial map exhibiting nontrivial behavior~\cite{friedland1989dynamical}.
Note that the parameter $a$ controls the degree of nonlinearity: for $a \gg 1$ the complete horseshoe is realized~\cite{devaney1979shift,bedford2004real}, while for $a \simeq 1$ KAM curves appear and the phase space becomes a mixture of regular and chaotic regions. 
It is easy to show that the H\'enon map can be converted to the map~(\ref{eq:cmap}) with a cubic potential~\cite{shudo2011complex}.

Since the H\'enon map is a polynomial map, it can be straightforwardly extended from $\H:{\Bbb R}^2 \mapsto {\Bbb R}^2$ to $\H:{\Bbb C}^2 \mapsto {\Bbb C}^2$~\cite{smillie1997complex}. 
The most important classification of the orbits is whether they remain bounded or diverge to infinity over time.  We introduce the sets, 
\begin{align}
&I^{\pm} := 
\{(x,y) \in \bC^2 | \lim_{n\to\infty} \H^{\pm n}(x,y) \to \infty \,\}, \\
&K^{\pm} :=  \{ (x,y) \in \bC^2  | \lim_{n\to\infty} \H^{\pm n}(x,y) ~{\rm is ~ bounded~ in}~\bC^2  \}.
\end{align}
Since the map is invertible, the dynamics can also be defined in the backward direction,
and one can similarly consider the asymptotic behavior of the dynamics under backward iteration.
We further define
\begin{eqnarray}
&&K = K^{+} \cap K^{-}, \\
&&J^{\pm} = \partial K^{\pm}, \\
&&J = J^+ \cap J^- .
\end{eqnarray}
Here $K$,~$J^{\pm}$ and $J$ are respectively called the filled Julia set, 
forward (resp. backward) Julia set and the Julia set~\cite{smillie1997complex}.

\subsection{Candidates for semiclassically contributing complex trajectories}
\label{sec:Julia_set}

Systematic studies of multidimensional complex dynamical systems were initiated in the 1990s by Bedford and Smillie~\cite{bedford1991,bedford1991b,bedford1992b,bedford1992a,bedford1998polynomial,bedford1998polynomial1,bedford1999polynomial,bedford2002polynomial}, as well as by Forn{\ae}ss and Sibony~\cite{fornaess1994complex,fornaess1995complex}.
Here, the results presented below are basically due to Bedford and Smillie.

The most powerful machinery for investigating complex dynamics in multiple dimensions is the so-called 
pluripotential theory (potential theory in several complex variables). 
Potential-theoretic approaches were already employed in one-dimensional dynamics by Brolin~\cite{brolin1965invariant}. However, while the complex function theory required there is the theory of a single complex variable, potential-theoretic arguments for two-dimensional systems make extensive use of tools from several complex variables. 
As a result, the required mathematical technicalities are highly involved, 
and we therefore omit derivations and background prerequisites.
For a detailed exposition of two-dimensional complex dynamics, we refer the reader to Ref.~\cite{morosawa2000}.
Readers interested in the connection between these results and tunneling effects should consult Ref.~\cite{shudo2011complex}.

The argument for two-dimensional complex dynamics begins by introducing the Green function defined by 
\begin{align}
\label{green3}
G^{\pm}(x, y)\equiv \lim_{n\to
+\infty}\frac{1}{2^n}
\log^+ \left\Vert \H^{\pm n} (x, y)
\right\Vert, 
\end{align}
where $\log^+t  := \max\{0, \log t\}$. 
Using the so-called convergence theorem of currents~\cite{bedford1991b},
one can construct a unique invariant measure $\mu=\mu^{+}\wedge\mu^{-}$, 
where $\mu^{\pm}$ are supports of $J^{\pm}$. 
In particular, the measure $\mu$ satisfies the following properties, implying a remarkable connection between the forward (resp.\ backward) Julia set and the stable and unstable manifolds:

\bn
{\bf Theorem (Bedford-Smillie \cite{bedford1991,bedford1991b,bedford1992a,bedford1992b}).}

{\it 

\begin{enumerate}

\item 
The measure $\mu$ is mixing and hyperbolic. 

\vspace{-0mm}
\item 
For any unstable periodic orbit $\mathfrak{p}$, 
$\displaystyle \overline{W^s(\mathfrak{p})} = J^+$ and ~$\overline{W^u(\mathfrak{p})} = J^-$ hold.

\end{enumerate}

\n
Here $W^s(\mathfrak{p})$ and $W^u(\mathfrak{p})$ denote the stable and unstable manifolds associated with the unstable periodic orbit $\mathfrak{p}$. The measure $\mu$ is said to be hyperbolic if the Lyapunov exponents  $\Lambda_1$ and $\Lambda_2$ associated with $\mu$ satisfy $\Lambda_1 >0> \Lambda_2$. 
}

\bigskip

From the first statement, we can say that the system is chaotic in the set supporting the invariant measure $\mu$. 
Ergodicity also follows immediately from the first statement. 
Notice that the theorem does not specify the condition for the nonlinear parameter $a$ in the H\'enon map $\H$, which means that the statement remains valid even when regular and chaotic orbits coexist in real phase space. 
All these properties hold only in the uniformly hyperbolic regime when one restricts the dynamics to the real plane~\cite{devaney1979shift,bedford2004real,arai2007hyperbolic}.

We now introduce the set given by 
\begin{eqnarray*}
\label{laputa}
\h{2mm}
{\cal C} :=
\bigl\{ \, (p, q)\in \mathcal{M}_
{\infty} \,|\, {\rm Im} \, S_n(p, q) \
{\rm converges~absolutely~at} \ (p, q) \,\bigr\}, 
\end{eqnarray*}
where $\mathcal{M}_{\infty}$ is defined by 
\begin{eqnarray}
\mathcal{M}_{\infty} :=  \bigcup_
{\beta\in\mathbb{R}}
\lim_{n\to\infty} \mathcal{M}_n^{~\ast, \,\beta}. 
\end{eqnarray}
For the H\'enon map $\H$, we can rigorously show that $ J^+ \subset \overline{\cal C}
\subset K^+$ holds if the topological entropy on ${\mathbb{R}^2}$ is positive~\cite{shudo2009a,shudo2009b}. 
Here $\overline{X}$  denotes the closure of the set $X$. 
We now aim to associate the set of orbits with convergent ${\rm Im}\, S_n(p,q)$ with invariant sets in the dynamics.
However, if we fix the initial and final data $\alpha$ and $\beta$, the classical orbits contributing to the semiclassical sum depend explicitly on those values, which cannot be compatible with any invariant sets of the system.
We therefore resolve this difficulty by allowing $\alpha$ and $\beta$ to take arbitrary values.
%
%

The most relevant condition specifying the set ${\cal C}$ is the absolute convergence of ${\rm Im} \, S_n(q, p)$. 
When the imaginary part ${\rm Im} \,S_n$ of the action is not absolutely convergent,  
the two situations are possible:
${\rm Im} \,S_n \to +\infty$ or 
${\rm Im} \,S_n \to -\infty$. 
The former type of orbits is negligible, but the latter cannot be excluded in the sense of magnitude. 
However, the divergence of ${\rm Im} \,S_n \to -\infty$ is obviously unphysical, and those orbits should be excluded as a result of the Stokes phenomenon \cite{adachi1989numerical,shudo1996stokes,shudo2008stokes,shudo2016toward}. 

Although no rigorous proof has yet been established, there are many reasons to believe that $K^{\pm}=J^{\pm}$~\cite{shudo2009a,shudo2009b,shudo2011complex}.
If this is indeed the case, the above statement simplifies to $\overline{\mathcal{C}}=J^{+}$, 
and thus substantially restricts the set of candidate orbits, indicating that the contributing complex orbits should at least be sought within $J^+$. 
Note, however, that the semiclassical dominant orbits are not yet sharply specified because $J^{+}$ still contains exponentially many orbits~\cite{bedford1991,bedford1991b,bedford1992b,bedford1992a}. 
Even within $J^{+}$, the imaginary part of the action, which mainly governs the magnitude of the contributions, can differ even among orbits in $J^{+}$.
Thefore, to single out the dominant orbit(s) in the semiclassical sum, one needs more detailed information on the underlying dynamics.

As the second narrowing-down process, 
the second statement of the above theorem provides an essential insight.
If $K^+ = J^+$ further holds, $\overline{\cal C} = \overline{W^s(\mathfrak{p})}$ follows. 
This does not imply ${\cal C} = W^s(\mathfrak{p})$, it nevertheless indicates that the set ${\cal C}$ can be well approximated by $W^s(\mathfrak{p})$. 
In addition, infinitely many unstable periodic orbits exist in real chaotic regions in mixed phase space, and the stable manifold $W^s(\mathfrak{p})$ of any $\mathfrak{p}$ runs close to the orbits in ${\cal C}$. Hence, the set ${\cal C}$ must contain infinitely many orbits, each of which is associated with the stable manifold $W^s(\mathfrak{p})$.

\subsection{Complexified stable and unstable manifold}
\label{sec:Complexified_stable_unstable}

As already mentioned, several works have pointed out the importance of stable and unstable manifolds in describing tunneling phenomena. 
As shown in Ref.~\cite{creagh1999homoclinic}, the optimal complex path, connected with quasi-real homoclinic orbits that explore the chaotic region, accounts well for the fluctuations of the tunneling splitting. This implies that stable and unstable manifolds play important roles in the description of tunneling-related observables.
Also, it was pointed out in Refs.~\cite{takahashi2000complex,takahashi2003complex,takahashi2005intrinsic,takahashi2006anomalously,levkov2007complex,levkov2007unstable,levkov2009signatures}, complex stable and unstable manifolds, obtained by extending the stable and unstable manifolds into the complex domain, become crucial when the strength of perturbation to an integrable limit exceeds a certain threshold. 

These studies concern stable and unstable manifolds that are asymptotic to a single fixed point in the phase space. 
On the other hand, Bedford and Smillie proved that ``for any unstable periodic orbit $\mathfrak{p}$ one has $\overline{W^s(\mathfrak{p})} = J^+$". 
At first glance, this theorem may seem to state only something simple, but its implications are significant.
To see this, let us consider the stable manifolds $W^{s}(\mathfrak{p})$ and
$W^{s}(\mathfrak{p}')$ associated with two distinct unstable periodic orbits
$\mathfrak{p}$ and $\mathfrak{p}'$. If $\mathfrak{p}\neq \mathfrak{p}'$, then
$W^{s}(\mathfrak{p})$ and $W^{s}(\mathfrak{p}')$ are obviously different invariant sets.
Nevertheless, by the theorem above we have
$
\overline{W^{s}(\mathfrak{p})}=\overline{W^{s}(\mathfrak{p}')}=J^{+},
$
which means that $W^{s}(\mathfrak{p}')$ necessarily runs through in an arbitrary neighborhood of $W^{s}(\mathfrak{p})$.
We emphasize that this does not contradict the fact that $\mathfrak{p}$ and $\mathfrak{p}'$ can be well
separated in the (real and complex) phase space.

In a non-integrable system, no matter how small the chaotic
region is (for instance, even in the ultra-near-integrable system discussed in
Subsec.~\ref{sec:ultra_integrable}), we can expect the existence of infinitely many unstable periodic orbits in 
real and complex phase spaces. Consequently, infinitely many stable
manifolds pass arbitrarily close to any given stable manifold
$W^{s}(\mathfrak{p})$.
Such a situation is realized in the real plane only when the system is
uniformly hyperbolic. 
What is remarkable is that this property always holds even
when chaotic and regular regions coexist in the real dynamics. 
This feature is reminiscent of riddled basins or the so-called Lakes of Wada~\cite{alexander1992riddled,kennedy1991basins,nusse1996basins,aguirre2009fractal}.

Furthermore, one can generally expect infinitely many periodic orbits to exist in the real phase space, whether in a chaotic region or in a small stochastic layer sandwiched between KAM curves inside a regular region. It then follows that, for instance, when moving from a regular region into the chaotic sea---or even when moving within a regular region bounded by KAM curves---the two regions are connected by complex trajectories guided by stable manifolds.

Another important point to emphasize is that trajectories guided by stable manifolds, i.e., trajectories on $J^+$, exhibit an {\it amphibious} nature. 
A trajectory that follows a stable manifold moves regularly within the (complex) regular region, but once the orbit enters the chaotic region, it behaves as a chaotic orbit. 
Therefore, complex trajectories in the Julia set naturally realize the composite object proposed in \cite{creagh1999homoclinic}: a combination of an optimal tunneling path and real homoclinic orbits in the real plane that are asymptotic to it.
This amphibious nature of the complex trajectories leads to amphibious eigenstates~\cite{hufnagel2002eigenstates,backer2005flooding,ishikawa2009recovery,ishikawa2010dynamical}.

\subsection{Complex orbits minimizing the imaginary part of the action}
\label{sec:Most_dominant}

From the properties of the complex stable and unstable manifolds described above, it follows that the complex trajectories minimizing the imaginary part of the action are those that approach the real plane along the complex stable manifold.

To explain this, let us consider a periodic orbit $\mathfrak{p}\in\mathbb{R}^2$ and suppose that the associated complexified stable manifold $W^{s}(\mathfrak{p})$ intersects the initial manifold ${\cal A}$ of the semiclassical propagator at some point, say $z(\mathfrak{p})\in W^{s}(\mathfrak{p})\cap{\cal A}$.
Recall that, in general, intersections between two locally two-dimensional manifolds
are isolated points in a four-dimensional space.
Now consider the points contained in a small neighborhood of $z(\mathfrak{p})$ in the initial manifold ${\cal A}$,
and observe how the orbits starting from this neighborhood evolve in time.
Since the central point $z(\mathfrak{p})$ lies on the stable manifold $W^{s}(\mathfrak{p})$,
it approaches the periodic point $\mathfrak{p}$ under forward iteration.

Notice that there should exist a one-dimensional curve in the neighborhood of $z(\mathfrak{p})$ in the initial manifold ${\cal A}$ such that the points on this curve, say ${\cal L}$, also come close to $\mathfrak{p}$ and then move along the unstable manifold of $\mathfrak{p}$ in the real plane, i.e., $W^{u}(\mathfrak{p})\cap\mathbb{R}^2$.
In other words, a point on the curve ${\cal L}$ taken sufficiently close to $z(\mathfrak{p})$
moves along $W^{u}(\mathfrak{p})\cap\mathbb{R}^2$ for large $n$.
Obviously, the closeness of the initial point to $z(\mathfrak{p})$ controls
how closely the corresponding orbit approaches the real plane after passing by $\mathfrak{p}$.
Therefore, after approaching $\mathfrak{p}$ along the stable manifold $W^{s}(\mathfrak{p})$,
such orbits do not accumulate Im\,$S$, because they behave almost like real orbits.
Since the final point of the semiclassical propagator, say $q$, must be real-valued,
orbits that first approach the periodic orbit $\mathfrak{p}$ and then evolve close to the real plane
are expected to contribute to the final semiclassical sum.
This argument implies that, to evaluate the tunneling probability from an initial state to another region, it suffices to evaluate the imaginary parts of the actions of complex trajectories that follow the stable manifolds of periodic orbits lying in the real plane.

\begin{figure}[htbp]
\centering
\hspace*{-1.5cm}
\includegraphics[width=1.1\linewidth, trim=0mm 120mm 0mm 120mm, clip]{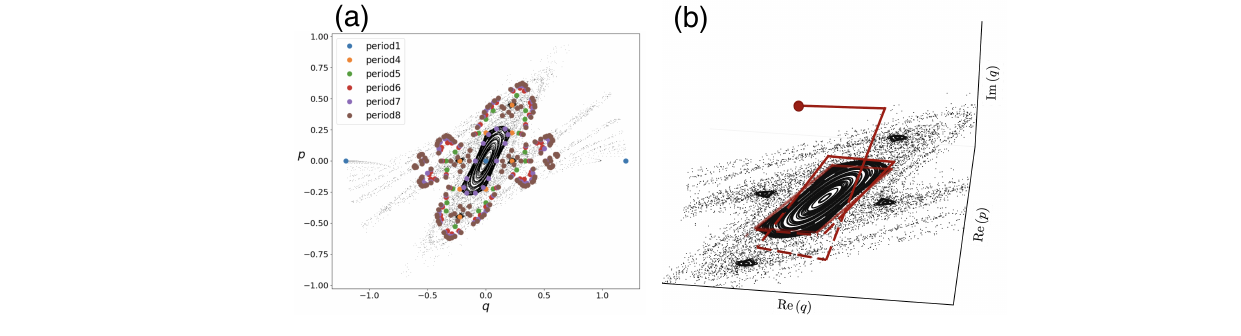}
\caption{
\label{fig:direct_path}
(a) Classical phase-space portrait and periodic orbits up to period 8 for the scattering map with potential $V(q)  = V_\kappa(q) + V_{\epsilon}(q)$ where $V_\kappa(q) = -\frac{\kappa}{16} \exp (-8q^2)$ and $V_\epsilon(q) =  - \epsilon \left[  {\rm erf}( \sqrt8  [q - q_b] ) - {\rm erf}(\sqrt8  [q + q_b]) \right]$. 
(b) Itinerary of an orbit 
tending to a period-4 periodic orbit indicated by the orange dots in (a).
For further details, see \cite{koda2022complexified}. 
}
\end{figure}

In Fig.~\ref{fig:direct_path}, we show an example of an orbit following the complex stable manifold that yields the smallest imaginary part of the action. 
The example considered here is the scattering map introduced in Ref.~\cite{mertig2018open}.
Figure~\ref{fig:direct_path}(a) displays the phase space of the scattering map together with periodic orbits obtained using the numerical techniques developed in Refs.~\cite{davidchack1999efficient,schmelcher1997detecting}; 
all periodic orbits up to period 8 were identified exhaustively.
We have computed the complexified stable manifolds emanating from each periodic orbit in the real chaotic region, and determined their intersections with the initial manifold ${\cal A}=I$, where $I$ denotes the approximate action variable labeling a torus within the regular region.
We then evolve each intersection point forward in time and compute the imaginary part of the action accumulated until the trajectory reaches the corresponding periodic orbit in the real plane.

By comparing these values, one can identify the stable-manifold trajectory that minimizes the imaginary part of the action. 
Note that the complexified stable manifold exhibits stretching-and-folding dynamics in complex space, analogous to the real stable manifold, and consequently becomes intricately entangled.
As a result,  the stable manifold from a single periodic point intersects with the initial manifold $I=\mathrm{const}$ (infinitely) many times.
Among these, the intersection point encountered first reaches the vicinity of the periodic point earliest and therefore accumulates the smallest imaginary action, and the remaining intersections give negligible contributions to the semiclassical sum.

The same applies to trajectories starting from intersections with complexified stable manifolds of different periodic points: the longer a trajectory spends in complex space, the larger the accumulated imaginary part of the action.
By computing the complexified stable manifolds emanating from each periodic orbit up to period 8, which is shown in Fig.~\ref{fig:direct_path}(a), and comparing the resulting imaginary actions,
we identified the trajectory shown in Fig.~\ref{fig:direct_path}(b) as the one with the smallest imaginary action.
This trajectory lies on the complexified stable manifold emanating from a period-4 orbit in Fig.~\ref{fig:direct_path}(a).
It spends the shortest time in complex space and therefore accumulates the smallest imaginary action.

We note that there exist two distinct period-4 orbits. The orbit that minimizes the imaginary part of the action is the one located closest to the regular region.
This can be verified by parametrizing the initial angle variable as $\theta=\xi+i\eta$:
the minimizing orbit has a smaller value of $|\,\eta\,|$ than the other period-4 orbit. 
This means that the trajectory with the smallest imaginary action is the one that departs from the vicinity closest to the real plane and reaches a location closest to the regular region.
The same scenario also accounts for the enhancement in the ultra-near-integrable systems~\cite{koda2022complexified}. 


This result is consistent with the observation of Doron and Frischat et al. that, among the possible tunneling paths, the one passing through the beach state becomes the most dominant tunneling route~\cite{doron1995semiclassical,frischat1998dynamical}.
In addition, the scenario in which the orbit tends to a periodic orbit in the real plane and subsequently evolves along an unstable manifold while exploring the chaotic sea is similar to the instanton-homoclinic composite identified by Creagh and Whelan~\cite{creagh1999homoclinic}.

Recalling the Bedford--Smillie theorem, $\overline{W^{s}(\mathfrak{p})}=J^{+}$, we emphasize, however, that many other complexified stable manifolds pass arbitrarily close to the complexified stable manifold obtained here.
In short-time semiclassical calculations, these manifolds are not resolved, but for long times they are distinguished as different trajectories. 
The tree-like structure appearing in Fig.~\ref{fig:M-set}(b), namely, a fractal structure in which branches sprout from a single trunk and further branches grow from those branches, is a manifestation of this fact~\cite{shudo2009a,shudo2009b}.
Therefore, even in the regular region, the contribution is not carried by a single optimal trajectory. Instead, exponentially many complex trajectories pass through the region and share almost the same imaginary part of the action.

\section{Conclusions and Outlook}
\label{sec:conclutions}

It is often said that chaos enhances the tunneling probability. 
When making such a statement, it is important to specify what reference is being used for the comparison.
The CAT mechanism asserts that, when one compares (i) a situation in which a chaotic region intervenes between symmetrically located classical tori with (ii) a situation where chaos does not appear and  the tori are directly coupled, the former exhibits an enhanced tunneling probability relative to the latter.
The behavior of the wavefunctions in Figs.~\ref{fig:hump}(c) and \ref{fig:hump}(d) 
captures exactly such a difference.
In the case where chaos is present, the wavefunction amplitude at $q=0$ is exponentially larger than in the integrable case.
The enhancement of the tunneling probability follows naturally from what one would expect based on the wavefunction profiles of mixed systems. 
In fact, this scenario is also incorporated into the hybrid semiclassical theory of Doron and Frischat~\cite{doron1995semiclassical,frischat1998dynamical}. 
They have argued that the transport through the chaotic region follows real classical dynamics there, 
which in turn is the mechanism by which chaos enhances tunneling. 
Moreover, as explained in Sec.~\ref{sec:Complex_path}, 
full semiclassical analysis, combined with the theory of complex dynamical systems, predicts the following: even after leaving the regular region there exist trajectories that continue to run deep in the complex domain, whereas others, after entering the chaotic region, stay near the real plane and move along chaotic trajectories on the real plane. 
Since the latter have a smaller imaginary part of the action than the former, they dominate the tunneling probability. 

In contrast, RAT concerns tunneling associated with transport through the regular region of phase space. 
In the RAT scenario, the enhancement is attributed to classical nonlinear resonances encountered along 
the transport through the regular region. 
Although, as in the case of chaos, nonlinear resonances almost certainly influence tunneling transport in one way or another, careful discussion is still needed to clarify how this effect manifests itself.
As emphasized in Subsec.~\ref{sec:QR}, it is important to distinguish between spikes in the tunneling splitting that appear when a tunneling doublet undergoes a crossing with a third state and the enhancement of the tunneling probability that persists even away from avoided crossings, which we referred to as persistent enhancement. 
One should recall that the former phenomenon can occur even in integrable systems.
As for the latter, the numerical results are most plausibly explained by a quantum-resonance-based scenario.
In any case, a fully semiclassical analysis is essential to clarify what actually happens in transport through the regular region. 
This may be regarded as one of the most important open problems concerning tunneling phenomena in non-integrable systems.
Another issue that should be pointed out is that, even if quantum resonance underlies the persistent enhancement, it arises only in systems driven by a periodic external force.
Further studies are needed to clarify whether a similar enhancement of the tunneling probability can be found in autonomous systems as well. 

Tunneling through a KAM region is a genuinely quantum phenomenon with no counterpart in real classical dynamics.
A classical interpretation therefore requires an extension of classical mechanics into the complex domain.
Unlike one-dimensional instanton tunneling, dynamical tunneling generally involves an intricate phase space, where regular and chaotic regions coexist even in the real domain. 
Because this complexity is expected to carry over to the complex domain, constructing a fully complex semiclassical theory for generic mixed systems becomes highly challenging.

The ultra-near-integrable systems introduced in Subsec.~\ref{sec:ultra_integrable} may provide an ideal setting for this purpose, because they exhibit an enhancement of the tunneling probability even though non-integrability-induced invariant structures do not appear in the corresponding classical phase space.
In particular, as indicated  in Subsec.~\ref{sec:ultra_integrable}, quantum resonances generate the step structure observed in the tail of the wavefunctions, and also underlie the step structure in the $\hbar$-dependence of the tunneling splitting in mixed phase space situations. 
Given that generic systems such as the standard map exhibit extremely intricate phase-space structures,
ultra-near-integrable systems would serve as a promising testing ground for the development of a semiclassical theory elucidating the origin of the persistent enhancement.
Indeed, Ref.~\cite{koda2022complexified} has already presented an analysis based on a full complex semiclassical theory in the time domain, and showed that the enhancement of the tunneling tail relative to the integrable limit can be explained in terms of the complex orbits with the least imaginary part of the action. 
The origin of the step structure has also been explored in terms of interference among the real actions of complex trajectories. 
Clarifying the role of classical nonlinear resonances in this context remains an important problem for future work.

\begin{ack}[Acknowledgments]

The authors are grateful to 
Arnd B\"acker, 
Stephan Creagh, 
Yasutaka Hanada, 
Srihari Keshavamurthy, 
Roland Ketzmerick, 
Ryonosuke Koda, 
Kensuke S. Ikeda, 
Yutaka Ishii,
Normann Mertig, 
Amaury Mouchet, 
Peter Schlagheck,
Kin'ya Takahashi, 
Steven Tomsovic, 
Denis Ullmo, 
and 
Diego Wisniacki
for their valuable comments and stimulating discussions. 
This work has been supported by JSPS KAKENHI Grant Numbers 
23K22417, 17K05583
and
25K07154 
\end{ack}

\seealso{article title article title}

\bibliographystyle{JHEP}%
\bibliography{reference.bib}

\end{document}